\documentclass[useAMS]{mn2e}
\usepackage{times}
\usepackage{psfig}

\def\gsim{ \lower .75ex \hbox{$\sim$} \llap{\raise .27ex \hbox{$>$}} } 
\def\lsim{ \lower .75ex\hbox{$\sim$} \llap{\raise .27ex \hbox{$<$}} } 
\def\ghi{$E_{\rm peak}-E_{\gamma}$}

\def\ama{$E_{\rm peak}-E_{\rm iso}$}
\def\yon{$E_{\rm peak}-L_{\rm iso}$}
\def\yonf{$E^{\rm obs}_{\rm peak}$--$P$}
\def\amaf{$E^{\rm obs}_{\rm peak}$--$F$}
\def\sw{{\it Swift}}
\def\sax{{\it Beppo}SAX}
\def\he{{\it Hete--II}}
\def\ko{{\it Konus--Wind}}
\def\ba{{\it BATSE}}
\def\ep{$E_{\rm peak}$}
\def\epo{$E^{\rm obs}_{\rm peak}$}
\def\epof{$E^{\rm obs}_{\rm peak}$--$F$}
\def\eiso{$E_{\rm iso}$}

\title[$E^{\rm obs}_{\rm peak}$ vs Fluence and Peak Flux]
{Peak energy of the prompt emission of long Gamma Ray Bursts vs their fluence
  and peak flux}

\author[Nava et al.]
{L. Nava$^{1,2}$\thanks{E--mail: lara.nava@brera.inaf.it}, G. Ghirlanda$^1$, 
G. Ghisellini$^1$ and C. Firmani$^{1,3}$\\ 
$^1$ Osservatorio Astronomico di Brera, via E.Bianchi 46, I-23807 Merate, Italy  \\ 
$^2$ Dipartimento di Fisica e Matematica, Universit\'a degli Studi dell'’Insubria, via Valleggio 11, I-22100 Como, Italy\\
$^3$ Instituto de Astronom\'{\i}a, Universidad Nacional Aut\'onoma de
M\'exico, A.P. 70-264, 04510, M\'exico D.F., M\'exico \\
}
\date{Accepted 2008 July 23. Received 2008 June 22; in original form 2008 April 11}

\begin{document}
   

\maketitle

\begin{abstract}
The spectral--energy (and luminosity) correlations in long Gamma Ray Bursts
are being hotly debated to establish, first of all, their reality against
possible selection effects. 
These are best studied in the observer planes, namely the
peak energy \epo\ vs the fluence $F$  or the peak flux $P$.
In a recent paper (Ghirlanda et al. 2008) we started to attack 
this problem considering all bursts with known resdhift and
  spectral properties.
Here we consider instead all bursts with known \epo,
irrespective of redshift, adding to those a sample
of 100 faint \ba\ bursts representative of a larger
population of 1000 objects.
This allows us to construct a complete, fluence limited,
sample, tailored to study the selection/instrumental effects we 
consider.
We found that the fainter \ba\ bursts have smaller \epo\ than
those of bright events. As a consequence, 
the \epo\ of these bursts is correlated with the fluence,
though with a slope flatter than that defined by bursts with $z$. 
Selection effects, which are present, are shown not to be 
responsible for the existence of such a correlation. 
About 6\% of these bursts are surely outliers of the \ama\
correlation (updated in this paper to include 83 bursts), 
since they are inconsistent with it for any redshift.
\epo\ correlates also with the peak flux,
with a slope similar to the \yon\ correlation.
In this case there is only one sure outlier.
The scatter of the \yonf\ correlation defined by the \ba\ bursts
of our sample is significantly smaller than the
\amaf\ correlation of the same bursts,
while for the bursts with known redshift the
\ama\ correlation is tighter than the \yon\ one.
Once a very large number of bursts with \epo\ and redshift will be available,
we thus expect that 
the \yon\ correlation will be similar to that currently found, whereas
it is very likely that the \ama\ correlation will become flatter and with a
larger scatter.

\end{abstract} 

\section{Introduction}

The correlation between the peak spectral energy \ep\ and the bolometric
isotropic energy \eiso\ emitted during the prompt (the so called "Amati"
correlation, Amati et al. 2002) may be a key ingredient for our comprehension
of the physics of Gamma Ray Bursts (GRBs).  The proposed interpretations
explain the \ama\ correlation as either due to geometric effects (Eichler \&
Levinson 2004; Toma, Yamazaky \& Nakamura 2005) or to the radiative process
responsible for the burst prompt emission (Thompson 2006; Thompson,
M\'esz\'aros \& Rees 2007), though there is no unanimous consensus.  In
addition, there is no agreement about the reality of the correlation itself.
Indeed, Nakar \& Piran (2005) and Band \& Preece (2005) have pointed out the
existence of outliers, while Ghirlanda et al.  (2005a) and Bosnjak et al.
(2008), considering an updated Amati relation, found no new outliers besides
GRB 980425 and GRB 031203.  More recently it has been argued that this
correlation might be the result of selection effects related to the detection
of GRBs (Butler et al.  2007 - hereafter B07).

The existence of an \ama\ correlation was predicted (Lloyd, Petrosian \&
Mallozzi 2000) before the finding of Amati et al. (2002). The prediction was
based on the existence of a significant correlation in the observer frame
between the peak energy \epo\ and the bolometric fluence $F$ of a sample of
\ba\ bursts. This finding has been recently confirmed by Sakamoto et al.
(2008a) using a sample of bursts detected by \sw, \ba\ and \he. In particular,
they note that X-Ray Flashes and X-Ray-Rich satisfy and extend this
correlation to lower fluences. Amati et al. (2002) discovered the \ama\ 
correlation with a sample of 12 bursts detected by \sax\ with
spectroscopically measured redshifts.  Later updates (e.g. Lamb, Donaghy \&
Graziani 2005; Amati et al.  2006; Ghirlanda et al. 2008 -- hereafter G08)
confirmed it with larger samples. In the most recent updates (76 bursts in G08
and 83 in this paper) the GRBs with measured redshift and \epo\ define a
correlation $E_{\rm peak}\propto E_{\rm iso}^{0.5}$, with only two outliers
(GRB 980425 and GRB 031203, but see Ghisellini et al. 2006).  This last sample
contains bursts detected by different instruments/satellites, i.e. \ba, \sax,
\he, \ko\ (all operative in the so--called pre--\sw\ era) and, since 2005
(mostly) by \sw.  These instruments have different detection capabilities and
also different operative energy ranges.

For these reasons is crucial to answer to the following question: is the \ama\ 
correlation real or is it an artifact of some selection/instrumental effect?

To investigate this issue we should move to the observer frame plane
corresponding to the Amati relation (i.e. the \epof\ plane) where the
instrumental selection effects act. There, we can place the ``cuts"
corresponding to the instrumental selection effects and see if the
distribution of the data points are affected by these cuts, or if, instead,
they prefer to lie in specific regions of the plane, away from these cuts.

There are two main selection effects: first, a burst must have a minimum flux
to be triggered by a given instrument. This minimum flux can be converted
(albeit approximately) to a fluence by adopting an average flux to fluence
conversion ratio, as done in G08. This is the minimum fluence a burst should
have to be detected.  We call it ``trigger threshold", TT for short.
Secondly, we need a minimum fluence also to find \epo\ and the spectral shape.
In fact we can have bursts that, although detectable, have too few photons
around \epo\ to reliably determine \epo\ itself.  Consider also that the
limited energy range of any detector inhibits the determination of \epo\ 
outside that range.  We call it ``spectral threshold" -- ST.

While both selection effects are functions of \epo\, we found that the 
latter is dominant for all the detectors. 

In G08 we considered a sample of 76 GRBs with redshifts $z$ (which we will
call, for simplicity, $z$GRB sample).  These bursts define an Amati
correlation in the form $E_{\rm peak}\propto E_{\rm iso}^{0.47\pm 0.03}$,
consistent with what found by previous works.  Moreover, G08 found that these
bursts define a strong correlation also in the observer frame (\epo$\propto
F^{0.39\pm0.05}$).

We demonstrated that the ST truncation effect is biasing the $z$GRB sample of
\sw--bursts (i.e. bursts for which \epo\ has been determined from the fit of
BAT spectra) while this is not the case for the no--\sw\ $z$GRBs (i.e.
  bursts for which \epo\ has been determined from other instruments, namely \ko,
  \ba, \sax\ and \he) . This leaves open two possibilities: (i) if those
described above are all the possible conceivable selection effects, the
no--\sw\ $z$GRB sample represents an unbiased sample and, therefore, the
\epof\ correlation defined with these bursts is real;
(ii) there are other selection effects biasing the sample of $z$GRBs.  For
instance, the optical afterglow luminosity might be proportional to the burst
fluence, resulting in a bias in favour of $\gamma$--ray bright bursts. Another
effect concerns the \ba\ bursts, that had to be localized by the Wide Field
Camera (WFC) of {\it Beppo}SAX.  Although, formally, the TT for the WFC should
not introduce any relevant truncation, in G08 we have shown that all bursts
detected by the WFC (with and without redshift or measurable \epo) had
fluences much larger than its TT curve.

To proceed further, in this paper we consider all bursts with measured \epo,
irrespective of having or not a measured redshift.

With this enlarged sample we study if the distribution of GRBs in the \amaf\ 
plane is: (i) consistent or not with the correlation defined by the $z$GRBs;
and (ii) if it is strongly biased or not by the considered selection effects.

Besides using existing samples of bursts (from \he, Sakamoto et
al. 2005, from \sw, B07, and from \ba, Kaneko et al. 2006, K06), we
collected a \ko\ sample from the GCN circulars (Golenetskii et
al., 2005, 2006, 2007, 2008) and we analysed a new sample of \ba\
bursts reaching a fluence of $10^{-6}$ [erg cm$^{-2}$].  This is the
\ba\  limiting fluence (ST) which allows to derive a reliable
\epo\ from the spectral analysis.  We therefore have a complete,
fluence limited, sample of \ba\ bursts.  With this we can study if
there is any \epof\ correlation and if it is a result of selection
effects or not.

By the same token, we can study the correlation between \ep\ and the
peak luminosity $L_{\rm iso}$.  This correlation was first found by
Yonetoku et al. (2004), with a small number (16) of bursts, and was
slightly tighter than the \ama\ correlation, and had the same slope.
It will be interesting to see if this is still the case considering
our sample of 83 $z$GRBs.  We can then investigate the instrumental
selection effects acting on this correlation by studying it in the
\yonf\ plane, where $P$ is the peak flux.

\begin{figure*}
  \vskip -0.5 true cm
  \centerline{\psfig{figure=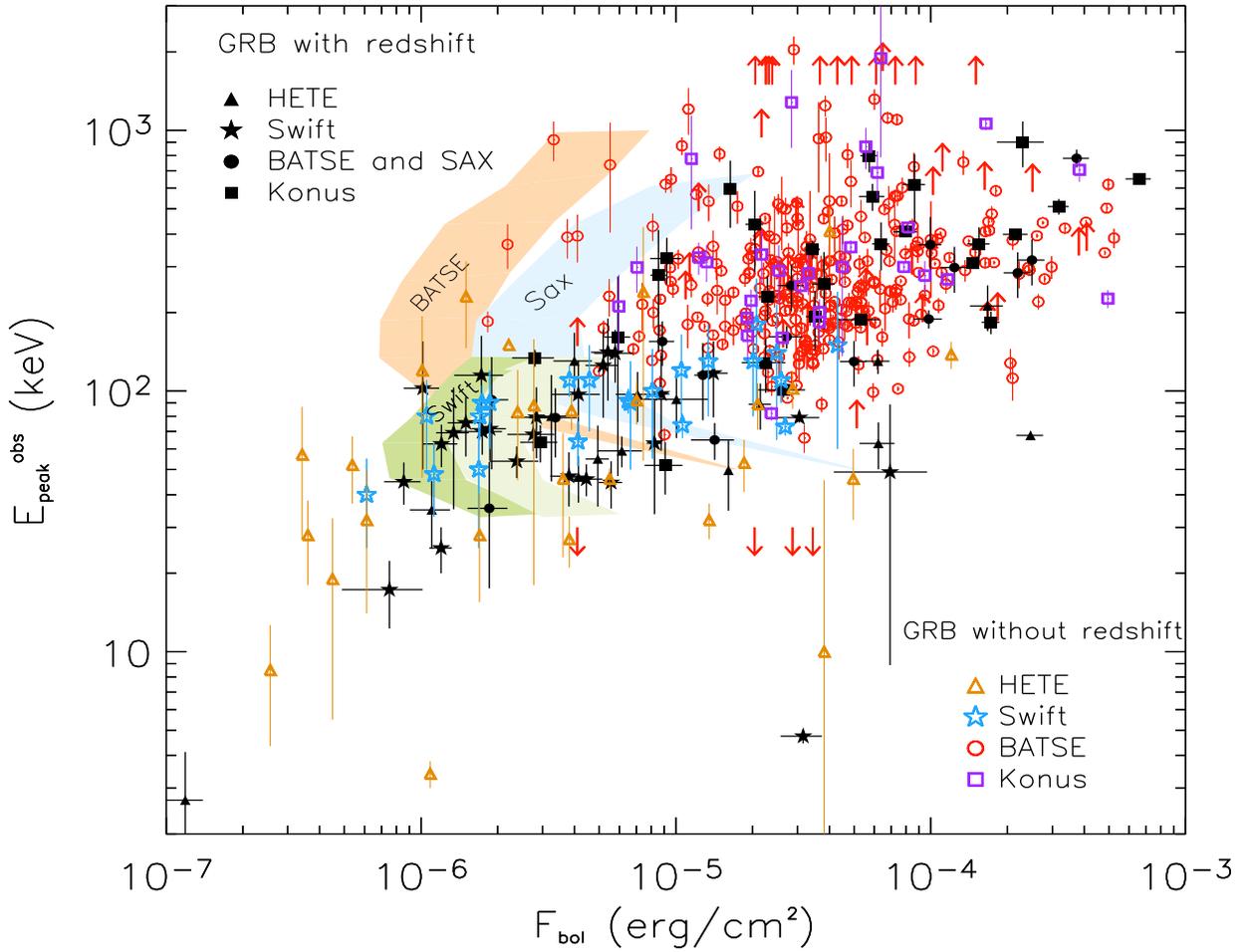,width=18cm,height=14cm}}
  \vskip -0.5 true cm
\caption{
  Distribution in the \amaf\ plane of the GRBs with measured redshift (filled
  symbols) and bursts without measured $z$ published in the literature (open
  symbols). The bolometric fluence is obtained by integrating the spectrum in
  the range 1 keV--10 MeV.  The bright \ba\ sample (from K06) is shown by open
  circles (for well constrained \epo) and up/down arrows (when only an
  upper/lower limit can be set on \epo\ from the spectral analysis of K06).
  \he\ (Sakamoto et al. 2005) and \sw\ bursts (B07 -- but see text) without
  redshifts are shown with open triangles and stars, respectively.  Shaded
  regions represent the ST curves of minimum fluence, for different
  instruments (see G08 for more details), down to which it is possible to fit
  the spectrum and constrain the spectral parameters.}
\label{fl_ep_kaneko}
\end{figure*}

\section{The \epo--Fluence plane}

The observer frame \amaf\ correlation, found using $z$GRBs (filled symbols in
Fig. \ref{fl_ep_kaneko}), divides the plane in two regions corresponding to
large fluence and low/moderate \epo\ (region--I) and low fluence and
large/moderate \epo\ (region--II).  The absence of bursts in region--I suggests
that they are extremely rare because, otherwise, they would have been easily
detected by present and past GRB detectors.  The absence of bursts in
region--II could be due to selection effects.

In the following we will refer to GRBs with known $z$ as $z$GRBs and we will
indicate bursts without measured $z$ as GRBs.

\subsection{GRB samples with redshifts: $z$GRBs}

Different detectors/satellites (\ba--CGRO, \he, \ko, \sax, \sw) have been
contributing the sample of GRBs with measured $z$ and \epo, possibly
introducing different instrumental selection effects. G08, by considering the
trigger efficiency of these satellites (from Band 2003), excluded that this is
affecting the sample of 76 bursts defining the \amaf\ correlation. However, a
stronger selection effect was studied in G08: to add a point in the \ama\ 
correlation, in addition to detect it, we need to determine (through the
spectral analysis) the peak energy \epo.  This requires a minimum fluence.
G08 simulated several spectra of GRBs by assuming they are described by a Band
function. The values assumed for the low and high energy indeces are fixed to
the typical values $\alpha=-1$ and $\beta=-2.3$.  By varying the peak energy,
the fluence and the duration G08 derived the ``spectral analysis thresholds''
ST (shaded curves in Fig.~\ref{fl_ep_kaneko}) in the \amaf\ plane for \ba,
\sax\ and \sw. Details of the simulations are given in G08. These curves
show that the limiting fluence $F$ is a strong function of \epo.  A burst on
the right side of these curves has enough fluence to constrain its peak
spectral energy.  As discussed in G08, Fig. \ref{fl_ep_kaneko} shows that the
27 \sw\ bursts (filled stars) with $z$ and well constrained \epo\ (from C07)
are affected by this selection effect (dark and light grey areas labeled \sw\ 
in Fig. \ref{fl_ep_kaneko}).  Note that this is a small sub--sample of the
bursts detected by \sw.  Indeed, in order to add a point to the \ama\ 
correlation, in addition to $z$, also \epo\ is needed. C07, through the
analysis of the BAT--\sw\ spectra of bursts with known $z$, showed that, given
the limited energy range of this instrument (15--150 keV), the peak energy
could be determined only for a small fraction of bursts.  Therefore, these 27
\sw\ bursts are all the GRBs (up to April 2008) for which both the $z$ is
known and \epo\ could be determined from the fit of the BAT--\sw\ spectrum.
The fact that the \sw\ sample, for which the spectral analysis of C07 yielded
\epo, extends to the estimated limiting ST is an independent confirmation of
the reliability of the method for simulating these curves.  Note that a few
\sw\ bursts are below these lines, but in these cases the peak energy was
found using the {\it combined} XRT--BAT spectrum (see G08).

Instead, pre--\sw\ $z$GRBs (partly detected by \ba\ and \sax\ -- filled
circles in Fig. \ref{fl_ep_kaneko}) are not affected by the corresponding ST
curves.  
Note that only for \ba\ we could derive the ST through our
simulations. To this aim, the detector response and background model is
needed. For \ko, \sax\ and \he\ these informations are not public. However,
for \sax\ we can rescale the \ba\ thresholds (see G08 for details). 

The sample of $z$GRBs considered in this work contains 83 objects, 76 from the
sample collected in G08, plus 7 bursts recently detected (up to April 2008).
For all these 7 bursts the spectral parameters come from fitting the 
{\it Konus--Wind} spectra and are reported in the GCN circulars (see Tab.
\ref{tabz}). With this updated sample we find an Amati correlation with slope
$s=0.48\pm 0.03$ and scatter $\sigma=0.23$\footnote{The scatter is found
  constructing the distribution of the logarithmic distance orthogonal
  to the best fit correlation line, and fitting this distribution with a
  Gaussian.  }.  The same sample defines also a correlation (Kendall's
$\tau=0.51$) in the observer plane in the form $E^{\rm obs}_{\rm peak}\propto
F^{0.40\pm0.05}$.

A way to investigate if the lack of GRBs in region--II of the \amaf\ plane,
i.e. between the distribution of bursts with $z$ and the ST curves defined in
G08, is real or it is due to a still unexplained selection effect is to
consider all GRBs with well constrained \epo\ but without measured $z$.

\subsection{GRB samples without $z$}

We consider \he\ bursts (Sakamoto et al. 2005), \sw\ bursts (from B07), the
bright \ba\ sample (K06) and \ko\ bursts (Golenetskii et al. 2005, 2006,
  2007, 2008, GCN circulars).  Through these samples we populate the \epof\ 
plane.

\subsubsection{Bright {\it BATSE} bursts}

We have considered the sample of 350 {\it BATSE} GRBs published by K06.
The selected bursts have either a peak photon flux (in the 50--300 keV energy
range) larger than 10 [photon s$^{-1}$ cm$^{-2}$] or a fluence (integrated
above 25 keV) larger than 2$\times 10 ^{-5}$ [erg cm$^{-2}$].  We excluded the
17 events whose spectrum was accumulated for less than 2 sec as most likely
representative of the short duration burst population.  With the remaining
GRBs, we constructed a first sample selecting all bursts whose time integrated
spectrum is fitted with a curved model (Band, cutoff power--law or
smoothly--joined power law) providing an estimate of \epo.  This sample
contains 279 GRBs.  The remaining bursts form another sample providing
lower/upper limits on \epo: those GRBs fitted with a Band or smoothly--joined
power law with an high energy photon index greater than --2 provide a lower
limit, as well as those fitted with a single flat power law.  On the contrary,
bursts fitted by single steep power laws (photon index $<-2$) provide an upper
limit on \epo.

Fig. \ref{fl_ep_kaneko} shows the K06 sample (open circles) in the \amaf\ 
plane together with the 83 $z$GRBs.  The \ba\ ST 
curves are also shown. By comparing \ba\ GRBs with $z$GRBs (filled symbols in
Fig. \ref{fl_ep_kaneko}) we note that the two samples are consistent for \epo\ 
values in the 100 keV -- 1 MeV range.  However, note that in the K06 sample
there are also a few bursts with considerably smaller fluence (but similar
\epo) with respect to $z$GRBs.  In other words, there is an indication of the
existence of bursts that lie between the limiting \ba\ curves and the \epof\ 
correlation defined by $z$GRBs (region--II). From Fig.~\ref{fl_ep_kaneko} it
is clear that the sample of bright \ba\ bursts is not strongly affected by the
corresponding ST. However, note that this sample is not appropriate
to study this issue because it is representative only of very bright \ba\ 
bursts and it is not complete in fluence.

The K06 sample shows a weak \epof\ correlation with a Kendall's correlation
coefficient $\tau=0.13$ (3$\sigma$ significance).

\subsubsection{\sw\ and \he\ bursts}

Other two samples of bursts with published spectral parameters are that of
\he\ and \sw. The two references for the \sw\ bursts are C07 and B07: the
former focused on \sw\ bursts with $z$ and the latter considered also bursts
without redshifts. The C07 \sw\ bursts were included in the sample of the 83
$z$GRBs. For the \sw\ bursts without $z$ we consider the analysis performed by
B07 (but see also Sakamoto et al. 2008b). They analysed GRB spectra with
either the frequentist method and through a Bayesian method. While the first
method allows to constrain the spectral \epo\ only if it lies in the energy
range where the spectral data are (15--150~keV for BAT--\sw), the bayesian
method infers the peak energy by assuming an \epo\ distribution as prior.  For
homogeneity with the analysis of C07 and with the method used to find the ST,
we consider only the \sw\ bursts of B07 without $z$ which have their peak
energy estimated through the frequentist method and for which this estimate
has a relative error $<$100\%.  This choice corresponds to the conditions of
the simulated ST of G08. We found 22 \sw\ bursts which satisfy these
requirements.

The {\it Hete--II} group published some spectral catalogs of their bursts
(Barraud et al. 2003; Atteia et al. 2005; Sakamoto et al. 2005). Sakamoto et
al. (2005) performed the time integrated spectral analysis of 45 GRBs detected
during the first 3 years of the {\it Hete--II} mission.  They performed
spectral fits by combining the data of the high energy detector (Fregate:
6--400 keV) and the low energy coded mask detector (WXM: 2--25 keV).  We have
considered in this sample the 27 bursts without $z$ (the remaining are already
included in the $z$GRB sample) and whose spectrum is fitted by a Band or
cutoff power--law model which provides an estimate of \epo.

Fig. \ref{fl_ep_kaneko} shows that \sw\ bursts (open stars) and \he\ bursts
(open triangles) are both consistent with the correlation defined in the
\amaf\ plane by the $z$GRB sample. Also the \sw\ sample without $z$ confirms
the validity of the ST estimates.  Note that the extension of \he\ events at
very low values of \epo\ is due to the fit of their spectrum with the WXM
instrument (see Sakamoto et al. 2005).

Note that in the sample of 83 bursts with $z$ there are also the {\it Beppo}SAX 
and {\it Konus--Wind} events.  No spectral catalog of bursts
without redshifts has been published to date for these two satellites.

\subsubsection{\ko\ bursts} 
Preliminary results arising from the fit of \ko\ spectra can be found in the
GCN circulars. We collected a sample of 29 GRBs (empty squares in
Fig.~\ref{fl_ep_kaneko}) for which an estimate of \epo\ and the spectral shape
is available and which are not already included in the \he\ or \sw\ samples
considered above. For each burst we estimate the bolometric (1--10$^4$ keV)
fluence and the bolometric peak flux. The results are listed in
Tab.~\ref{tabkonus} in the Appendix. Since \ko\ covers an energy range from 20
keV to a few MeV, a good spectral analysis of very hard bursts can be
performed.  Anyway, the determination of its TT and ST is not possible, as the
background model and the detector response are not public. Therefore, with
respect to the distribution of these GRBs in the \amaf\ plane
(Fig~\ref{fl_ep_kaneko}), we can only note that it seems to be very similar to
that of bright \ba\ bursts (empty circles).

\begin{figure*}
\vskip -0.5 true cm
\centerline{\psfig{figure=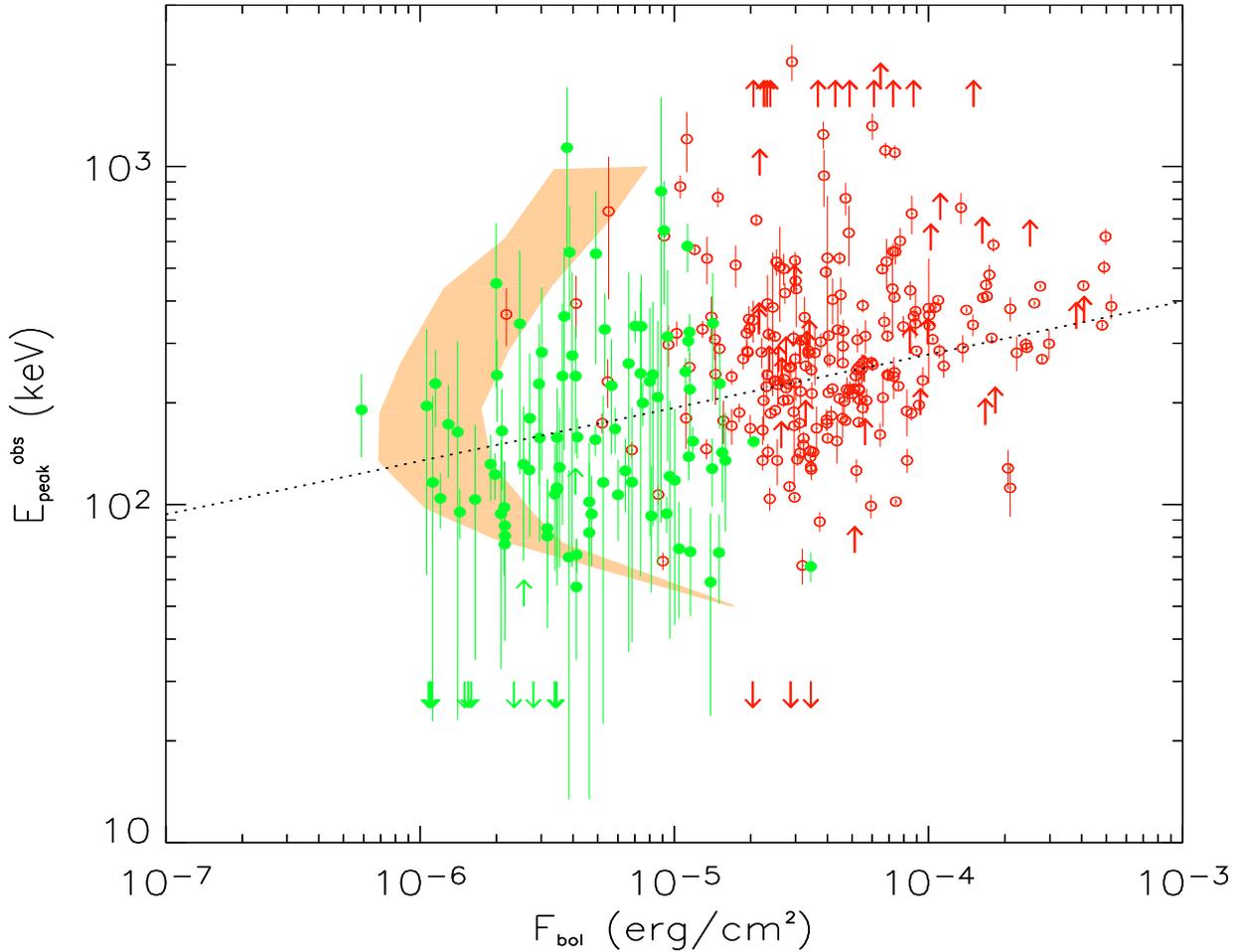,width=18.cm,height=14.cm}}
\vskip -0.2 true cm
\caption{The \amaf\ plane for the sample of \ba\ bursts. The fluence reported in this
  plot is the bolometric fluence (1--10$^4$ keV). The open circles are GRBs of
  the K06 sample with catalog fluence larger than 2$\times 10^{-5}$ erg
  cm$^{-2}$ and the filled circles are the 100 GRBs analyzed in this paper.
  The arrows correspond to those bursts for which we can only estimate a
  lower/upper limit to \epo. The shaded region represents the minimum fluence
  requested to constrain \epo\ from the spectral analysis. The left and right
  boundaries of this region are calculated for a burst lasting 5 and 20
  seconds respectively (see G08 for more details). The dotted line represents
  the best fit to the combined sample: $E^{\rm obs}_{\rm peak}\propto F_{\rm
    bol}^{0.16\pm 0.02}$.}
\label{solobatse}
\end{figure*}

\section{Faint {\it BATSE} bursts}

Among the above sample, the bright \ba\ bursts of K06 is the largest and,
given the spectral range of \ba, covers a wide range in \epo.  However, this
sample was selected according to a minimum peak flux {\it or} fluence
threshold and it is not complete either in peak flux and fluence.  Furthermore
it extends down to a fluence larger than the the minimum fluence required to
derive \epo.

Samples reaching smaller fluences indeed exist: Yonetoku et al.  (2004)
performed the spectral analysis of 745 GRBs from the \ba\ catalog with flux
larger than $2\times10^{-7}$ [erg cm$^{-2}$ s$^{-1}$].  However, they exclude 56
GRBs for which they find a pseudo redshift greater then 12 or no solution
using the $E_{\rm peak}- L_{\rm iso}$ relation as a distance indicator.  Thus
the final sample is biased by this choice as it most likely excludes the
bursts with low fluence and intermediate/high peak energy and therefore is not
representative of the whole sample of \ba\ bursts with fluences greater than
the spectral threshold.

It is clear from Fig. \ref{fl_ep_kaneko} that for \ba\ bursts ``there is
room'' to extend the K06 sample to smaller fluences: if the ST for
\ba\ are correct, we should be able to analyze bursts with fluence smaller
than 2$\times 10^{-5}$ [erg cm$^{-2}$].  Therefore, in order to increase the
statistics and test the density of bursts in region--II we extend the \ba\ 
sample to the limiting fluence of 10$^{-6}$ [erg cm$^{-2}$].  To this aim we
selected a representative sample of
100 \ba\ bursts with a fluence $F$ above 25 keV (which is a good proxy for the
bolometric fluence), within the range $10^{-6} <F< 2\times 10^{-5}$ [erg cm$^{-2}$].  
The number of extracted GRBs per fluence bin follows the
Log$N$--Log$F$ distribution and therefore this sub--sample is representative
of the \ba\ burst population in this fluence range (which corresponds to $\sim
1000$ events).

For all these bursts we analysed the {\it BATSE} Large Area Detector (LAD)
spectral data which consist of several spectra accumulated in consecutive time
bins before, during and after the burst.  The spectral analysis has been
performed with the software \emph{SOAR} v3.0 (Spectroscopic Oriented Analysis
Routines), which we implemented for our purposes.  For
each burst we analysed the {\it BATSE} spectrum accumulated over its total
duration (which in most cases corresponds to the $T_{90}$ parameter reported
in the {\it BATSE} catalog). In order to account for the possible time
variability of the background we modeled it as a function of time (see e.g.
K06).

In most cases we could fit either the Band model (Band et al. 1993) or a
cutoff power law model. To be consistent with the method used to derive the
spectral threshold curves of Fig. \ref{fl_ep_kaneko}, we consider that \epo\ 
is reliably determined if its relative error is less than 100\%.  If the
relative error is greater or if the best fit model is a simple power law we
derive the corresponding lower/upper limit.  In these cases the burst is
reported on the plot as an up/down arrow.

In Tab.~\ref{tabverdi} in the appendix we list the bursts of our sample
together with the results of the spectral fitting.

\begin{figure}
  \hskip -0.2 true cm
  \centerline{\psfig{figure=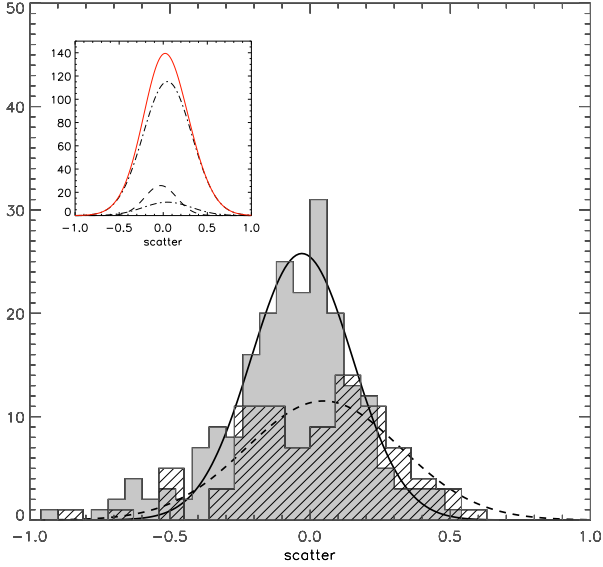,width=9cm,height=8cm}}
\caption{ Scatter distributions of the sample of \ba\ bursts with well
  estimated \epo\ around their best fit correlation in the \amaf\ plane (see
  Fig.~ \ref{solobatse}).  The shaded distribution is of the 213 GRBs of K06
  and the hatched distributions is of the 88 GRBs of our analysis (we have
  excluded upper/lower limits).  The fit with gaussian functions of the two
  distributions have scatter $\sigma=0.18$ for the K06 sample and
  $\sigma=0.29$ for our sample.  The insert shows the scatter of the combined
  sample ($\sigma=0.26$) (solid line), once we take into account that
  the GRBs analyzed by us are representative of 1000 bursts.}
\label{scat}
\end{figure}

\begin{figure}
 \hskip -0.2 true cm
\centerline{\psfig{figure=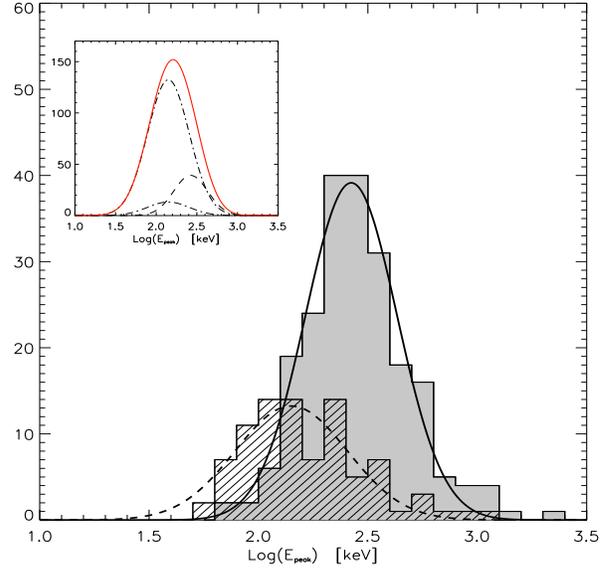,width=9.cm,height=8cm}}
\caption{ 
  The \epo\ distributions for the K06 GRB sample (213 GRBs -- shaded
  histogram) and for the sample that we have analyzed (88 GRBs with well
  determined \epo\ over 100 bursts randomly extracted from the \ba\ 
  Log$N$--Log$F$ distribution in the range $10^{-6} <F< 2\times 10^{-5}$
  [erg~cm$^{-2}$] -- hatched histogram).  The solid and dashed lines are the
  gaussian fits to the two distributions. The K--S test (see text) confirms
  that the shift of \epo\ to lower values for dimmer bursts is statistically
  significant. In the insert are reported these two gaussian fits and their
sum (solid line) which has been obtained by multiplying the distribution of
the 100 GRBs by 10 (dot--dashed line) in order to account for the total number
of bursts in this fluence range.}
\label{istobatse}
\end{figure}

\subsection{Results for the complete sample of \ba\ bursts}

In order to construct a complete sample of \ba\ bursts, we cut the K06 sample
at a fluence $F$ (as reported in the \ba\ catalog) greater than $2\times 10
^{-5}$ [erg cm$^{-2}$] (213 GRBs). This complete sub--sample is representative
of bright \ba\ bursts. To this we add the 100 bursts of our representative
sample of the 1000 GRBs with fluences between $ 10^{-6}$ and 2$\times 10
^{-5}$ [erg cm$^{-2}$].  Fig. \ref{solobatse} shows the \epo\ and bolometric
fluence (computed in the range 1 keV -- 10 MeV) of the sub--sample of bursts
from the K06 sample (open circles) together with the 100 bursts of our sample
(filled circles).  This combined sample extends the K06 fluence limit to
$F>10^{-6}$ erg cm$^{-2}$. Note that in this figure we plot the bolometric
(1-10$^4$ keV) fluence estimated accordingly with the best fit model. Its
value can be different from the fluence value reported in the \ba\ catalog.

The distribution of \ba\ GRBs in Fig. \ref{solobatse} defines a correlation
with a large scatter. The Kendall's correlation coefficient is $\tau=0.18$
(7$\sigma$ significant). Since the dimmer part of the burst distribution in
the \amaf\ plane is affected by the ST truncation effect, we analyzed the
correlation also following the method proposed by Lloyd et al. (2000). We
obtain a Kendall's correlation coefficient $\tau=0.2$ (5.5$\sigma$
significant). By fitting with the least square method, without weighting for
the errors and neglecting the upper/lower limits, we obtain $E^{\rm obs}_{\rm
  peak}\propto F_{\rm bol}^{0.16\pm 0.02}$ (dotted line in Fig.
\ref{solobatse}).

In Fig. \ref{scat} we show the distribution of the scatter of the two samples
around the best fit correlation in the \amaf\ plane.  These have standard
deviation of $\sigma$=0.18 for the K06 sample (solid histogram) and
$\sigma~$=0.29 for our representative sample (hatched histogram).  The
combined sample (solid line in the insert) has a scatter distribution with
$\sigma~$=0.26.  In order to describe the scatter distribution of \ba\ bursts
down to the fluence limit of $F>10^{-6}$ [erg cm$^{-2}$], we have to consider
that our sample of 100 bursts is representative of the entire burst population
(a factor 10 larger in number) in the fluence range $10^{-6} <F< 2\times
10^{-5}$ [erg cm$^{-2}$].  We fitted the scatter distributions of our (dashed
line) and K06 (solid line) sample with gaussian functions and combined these
best fit distributions by renormalizing that of our sample by a factor 10
(corresponding to the ratio of the extracted bursts with respect to the total
number of \ba\ bursts in the same fluence bin). The result is shown in the
insert of Fig. \ref{scat} (solid line).  The combined sample (solid line in
the insert) has a scatter distribution with $\sigma$=0.26.

Fig. \ref{solobatse} shows that there are bursts with low fluence and high
\epo\ and that the dispersion in \epo\ at low fluence is larger then the
dispersion at high fluence.  However, Fig.~\ref{solobatse} also shows that, on
average, the error on the \epo\ value increases for smaller fluences. This
could imply that the larger scatter for smaller fluences is in part due to
larger errors on \epo.  A simple way to determine the contributions to the
total observed scatter $\sigma_{\rm tot}$, calculated orthogonally to the
fitting line, is:
\begin{equation}
\sigma_{\rm tot}^2 \, =\, \sigma^2_E\cos^2\theta + \sigma^2_{\rm c} 
\end{equation}
where $\sigma^2_E$ is the average relative error on \epo, $\theta$ is the
angle defined by the slope of the correlation (whose angular coefficient is
equal to $\tan\theta=0.16$) and $\sigma_{\rm c}$ is the intrinsic scatter of
the distribution.  For fluences greater than 2$\times 10^{-5}$ [erg cm$^{-2}$]
(K06 sample) $\sigma_{\rm tot}\sim \sigma_{\rm c}=0.18$ since the errors on
\epo\ are small.  On the other hand, for fluences smaller than 2$\times
10^{-5}$ [erg cm$^{-2}$] (our sample), $\sigma_E=0.18$ and
$\sigma_{\rm tot}=0.29$, leading to $\sigma_{\rm c}=0.23$, to be compared to
$\sigma_{\rm c}=0.18$ for fluences greater than 2$\times 10^{-5}$ [erg
cm$^{-2}$].  This leads us to conclude that the intrinsic scatter around the
best fit line increases for smaller fluences.  A caveat is in order: the
scatter $\sigma_{\rm c}$ does not take into account lower/upper limits, which
also do not enter in the derivation of the best fit line.  Thus $\sigma_{\rm
  c}$ could be larger, but only slightly, since the number of upper/lower
limits is very limited.

Through our \ba\ sample we can also study the \epo\ distribution. In
Fig. \ref{istobatse} we show the \epo\ distribution of our \ba\ sample
(hatched histogram) and that of bright \ba\ bursts of K06 (solid filled
histogram -- cut at 2$\times 10^{-5}$ [erg cm$^{-2}$]).

The shift of the \epo\ distribution to lower values for smaller fluence
selection is statistically significant: the K--S test gives a probability
$P=5.8\times10^{-11}$ that the two distributions belong to the same
population.  Similarly to what has been done for the scatter distribution, we
combine the two distributions by accounting for the fact that our sample is
representative of a larger population of bursts in the $10^{-6}$-- 2$\times
10^{-5}$ [erg cm$^{-2}$] fluence range. The result is shown in the insert of
Fig. \ref{istobatse} (solid line). 
This total sample distribution has a peak at \epo$\sim$160 keV, 
i.e. smaller than the 260 keV of bright bursts of
K06, and a standard deviation $\sigma$=0.28.  
Although the widths of the distributions of \epo\ can be affected by the 
measurement errors, the central values are not.

\section{Outliers of the \ama\ correlation}

\begin{figure*}
\vskip -0.3 true cm
\hskip -0.3 true cm
\centerline{\psfig{figure=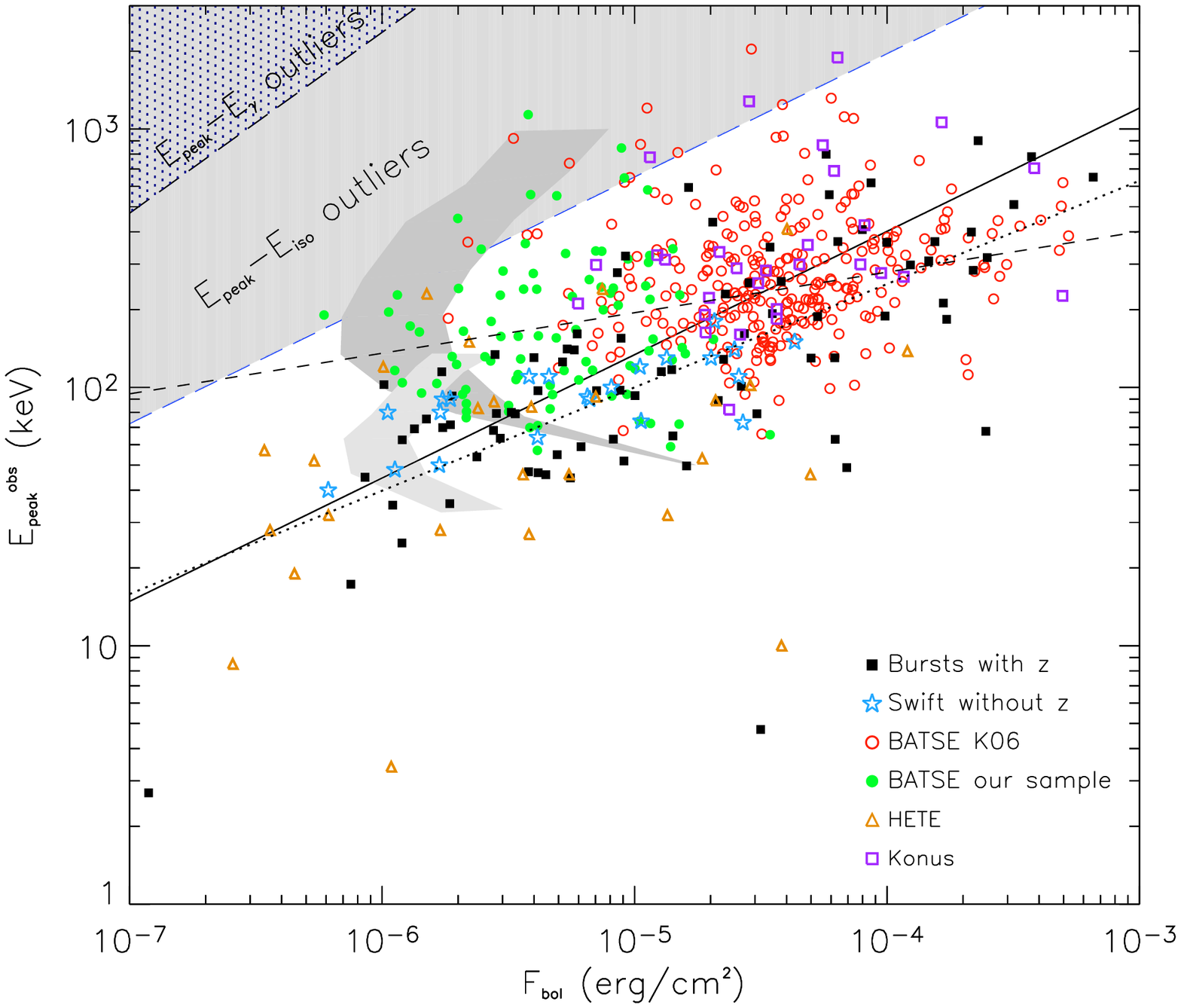,width=18cm,height=15cm}}
\vskip -0.3 true cm
\caption{
  Consistency test of the \ama\ correlation. The open symbols represent the
  bursts without redshifts detected by \he\ (triangles), \ba\ (open and filled
  circles), \sw\ (open stars) as described in the legend.  The solid line is
  the \ama\ correlation transformed in the observer \amaf\ plane. The
  long--dashed line is the 3$\sigma$ scatter of the \ama\ correlation.  The
  ``region of outliers" is the grey shaded.  Bursts falling in this region are
  not consistent with the \ama\ correlation for any redshift: they are
  outliers at more than $3\sigma$ (if the scatter distribution is Gaussian,
  see text). In the upper left corned we also show the ``region of outliers''
  of the \ghi\ correlation (adapted from Ghirlanda et al. 2007) if bursts have
  a 90$^\circ$ jet opening angle.  The dotted line is the fit to the $z$GRB
  sample (filled squares) and the dashed line is the fit to the complete
  sample of \ba\ bursts described in Sec.3.1 (see also Tab.\ref{tab1})}
\label{outliers_ama}
\end{figure*}

In Fig. \ref{outliers_ama} we combine, in the \epof\ plane, our sample
of \ba\ bursts with the $z$GRBs (solid filled squares) and with
the \sw, \he\ and \ko\ samples of GRBs without redshifts.

We note that bursts with known redshift (filled squares) are only
representative of the large fluence (for any \epo) part of the plane.
In particular, Fig. \ref{outliers_ama} shows the existence of bursts
with low fluence (between $F\leq 10^{-6}$ and $F\sim 10^{-5}$ [erg
cm$^{-2}$]) but \epo\ larger than 200 keV. These events are not
present in the $z$GRB sample.  Their absence in the $z$GRB sample
suggests the existence of a selection effect.

However, $z$GRBs are those defining the Amati correlation (as in G08). We do
not know if all the other bursts (without redshifts) represented in
Fig. \ref{outliers_ama} satisfy this correlation. 

%

If these GRBs have a similar redshift distribution of those with measured $z$,
then it is likely that they would define a rest frame \ama\ correlation with
different properties (slope, normalization and scatter), since some of them
stay apart from the \amaf\ correlation defined by the $z$GRB sample.  On the
other hand, GRB 980425 and GRB 031203 do have a peak energy and fluence
consistent with the $z$GRB sample, but it is their small redshift to make them
outliers with respect to the \ama\ correlation.

The possibility that there is a considerable number of outliers of the \ama\ 
correlation in the \ba\ sample has been discussed in the literature (e.g.
Nakar \& Piran 2005; Band \& Preece 2005; K06 -- but see also Ghirlanda et al.
2005; Bosnjak et al. 2007).  We can test if a burst is consistent or not with
the \ama\ correlation even if we do not know its redshift.  Simply, we assign
to the burst any redshift, checking if there is at least one $z$ making it
consistent with the correlation. By ``consistent" we mean the the burst must
fall within the 3$\sigma$ scatter (assumed gaussian) of the correlation.  This
test was first proposed for the short bursts (Ghirlanda et al.  2004a) and
then applied to \ba\ long GRBs.  More quantitatively, following Nakar \& Piran
(2005), we can write the \ama\ correlation as
\begin{eqnarray}
E_{\rm peak}^{\rm obs} (1+z) &=& k \left( { 4\pi d_{\rm L}^2 F \over 1+z}\right)^a 
\, \, \to \nonumber \\ 
E_{\rm peak}^{\rm obs} &=& k F^a f(z);  \qquad
f(z)={ (4\pi d_{\rm L}^2)^a \over (1+z)^{1+a} } 
\end{eqnarray}
where $F$ is the bolometric fluence.  The function $f(z)$ has a
maximum ($f_{\rm max}$) at some redshift and therefore all bursts for
which $E_{\rm peak}^{\rm obs}/(k F^a)>f_{\rm max}$ are outliers.  We
can impose that the constant $k$ accounts for the scatter of the best
fit \ama\ correlation, and then find outliers at some pre--assigned
number of $\sigma$.  It is worth to recall that this method assumes
that the dispersion of points, around the \ama\ correlation under
test, is described by a Gaussian function.  With this assumption we
can state that a given GRB is $N\sigma$ inconsistent with the
correlation, and quantify the probability of having a certain number
of outliers lying -- say -- more than 3$\sigma$ away.  Since the Amati
correlation, as discussed below, surely incorporates an
extra--Poissonian dispersion term (Amati, 2006), the scatter
distribution may not be a Gaussian, but it may correspond to the
distribution function of this extra term.  In other words: the scatter
of the points around the Amati correlation {\it is not} due to the
errors of our measurements, but reveals the presence of an
extra--observable not considered in the Amati relation.  With this
caveat, we nevertheless use this assumption (i.e. Gaussianity) for
simplicity.

In Fig. \ref{outliers_ama} the grey area identifies, in the \amaf\
plane, the ``region of outliers".  Considering only \ba\ bursts
we can state that the 6\% of the complete sample considered in this
paper is constituted by bursts which are surely outliers of the \ama\
relation.  We can test if these outliers have different spectral
properties with respect to other bursts (that instead pass the above
consistency test).  By comparing their spectral parameters we find
that the outliers of the \ama\ correlation have a larger peak energy
than the total sample of bursts (K--S probability
$P=8.7\times10^{-5}$) and a slightly harder low energy spectral index
$\alpha$ (K--S probability $P=10^{-1}$).  From Fig.~\ref{outliers_ama}
we also note that there is no outlier for the $E_{\rm
peak}$-$E_{\gamma}$ correlation.

\begin{table}
\begin{tabular}{|clccc|}
\hline
{\bf correlation}&{\bf sample}      & {\bf scatter}&{\bf K}&{\bf s} \\
\hline
\yon\            & $z$GRB           & 0.28         & --18.6    & 0.40$\pm$0.03 \\
\cline{2-5}
                 & $z$GRB (only \sw)&              & --11.4    & 0.26$\pm$0.05 \\
\cline{2-5}    
                 & $z$GRB (not \sw) &              & --20.5    & 0.44$\pm$0.03 \\
\hline
\ama\            & $z$GRB           & 0.23         & --22.7    & 0.48$\pm$0.03 \\
\cline{2-5}
                 & $z$GRB (only \sw)&              & --16.7    & 0.36$\pm$0.06 \\
\cline{2-5}
                 & $z$GRB (not \sw) &              & --24.4    & 0.51$\pm$0.03 \\
\hline
\yonf\           & $z$GRB           & 0.26         & 4.41      & 0.39$\pm$0.05 \\
\cline{2-5}
                 & \ba\             & 0.20         & 3.93      & 0.28$\pm$0.02 \\
\hline  
\amaf\           & $z$GRB           & 0.23         & 4.00      & 0.40$\pm$0.05 \\
\cline{2-5}          
                 & \ba\             & 0.23$^a$     & 3.07      & 0.16$\pm$0.02 \\ 
\hline
\end{tabular}
\caption{Results of the correlation analysis. For each correlation in the rest 
frame and observed plane we give the values of the scatter, normalization and slope. 
The correlations are in the form $y=K x^s$, where y is the logarithmic observed/rest 
frame peak energy  in units of keV and x is the logarithm of the luminosity (energy) 
in erg s$^{-1}$ (erg) or the peak flux (fluence) in erg cm$^{-2}$ s$^{-1}$
(erg cm$^{-2}$). $a$: this is the value once depurated of the contribution to
the scatter of the measurement errors (see text).}
\label{tab1}
\end{table}

\section{The \yon\ correlation}

\begin{figure*}
\vskip -0.5 true cm
\centerline{\psfig{figure=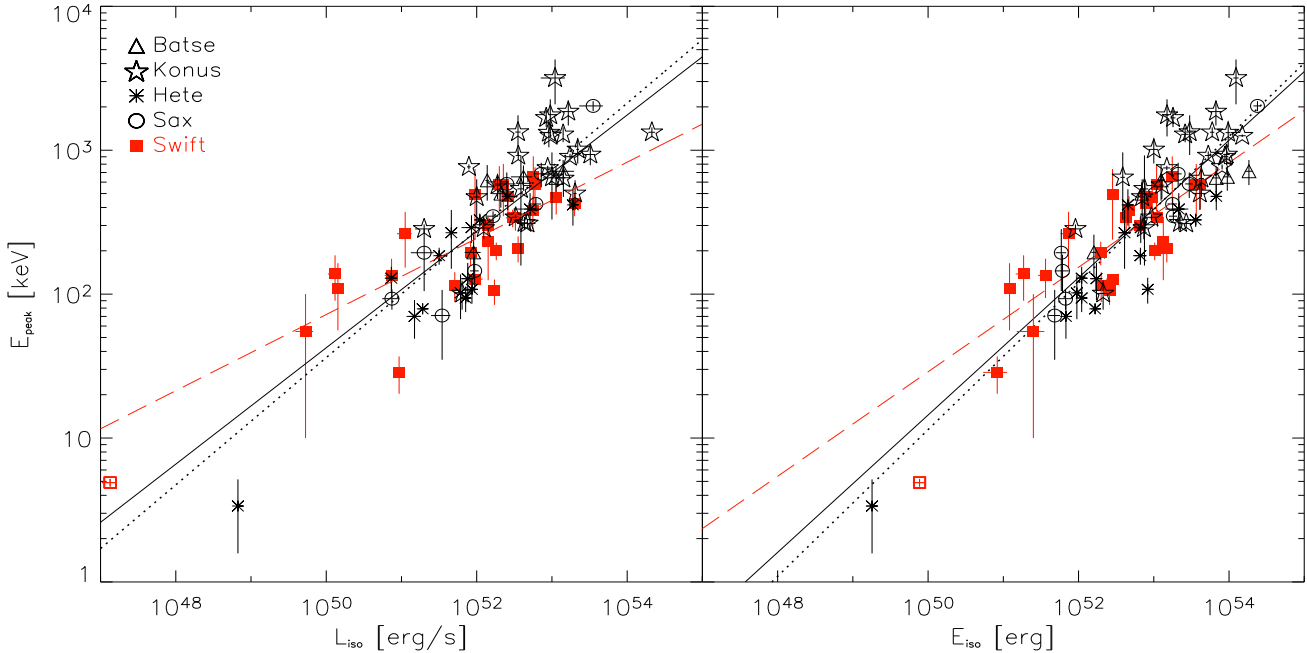,width=18.cm,height=9cm}}
\vskip -0.1 true cm
\caption{\yon\ and \ama\ correlations for 83 GRBs with measured redshift and
  spectral parameters. The best fit of the whole sample is shown with a solid
  line. Note that the fit performed on the \sw\ sample alone (filled squares)
  has in both cases a very flat slope (dashed line) with respect to the
  slope derived for no--\sw\ bursts (dotted line). The results of these
  different analysis are reported in Tab. \ref{tab1}. For an explaination of
  the flat slope found with the \sw\ sample see G08.}
\label{yone}
\end{figure*}
\begin{figure*}
\vskip -0.3 true cm
\centerline{\psfig{figure=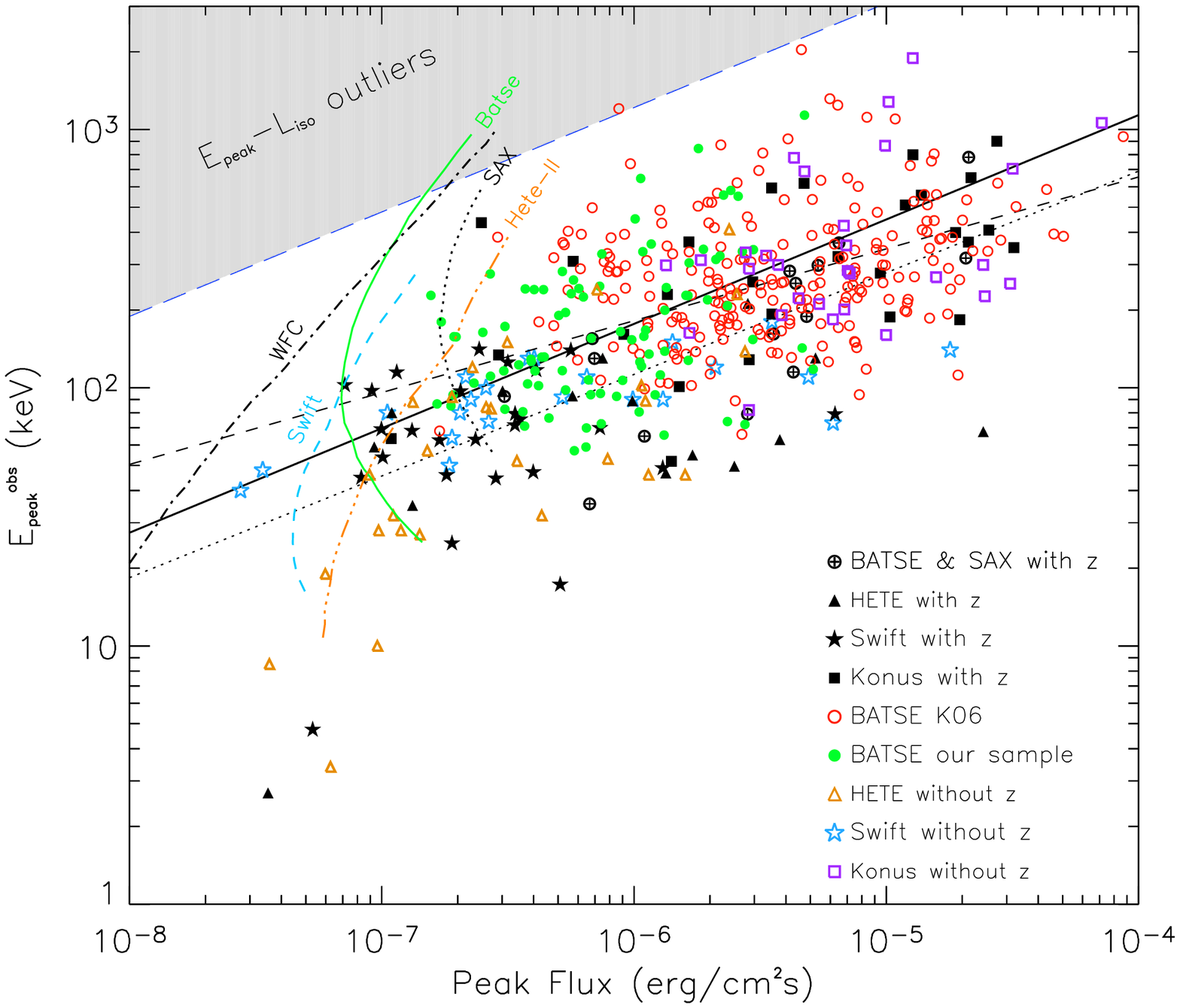,width=18.cm,height=15cm}}
\vskip -0.5 true cm
\caption{Consistency test of the \yon\ correlation. The filled squares are
  bursts with measured redshifts. The solid line is the \yon\ correlation
  transformed in the observer \yonf\ plane (here it is represented the
  bolometric peak flux). The long dashed line is the 3$\sigma$ scatter of
  the \yon\ correlation (as discussed in Sec. 5). The shaded triangle
  delimits the ``region of outliers".  Bursts falling in this region are not
  consistent with the \yon\ correlation for any redshift: they are outliers at
  more than $3\sigma$ (if the scatter distribution is Gaussian, see text). The
  dotted and dashed lines show the best fit obtained by considering
  respectively the $z$GRB sample and the \ba\ complete sample. The curves
  represent the TT estimated for different instruments by assuming that the
  trigger is based on the peak flux criterion.}
\label{outliers_yone}
\end{figure*}

Yonetoku et al. (2004, Y04), with a sample of 16
GRBs of known $z$, found that \ep$\propto L_{\rm iso}^{0.5}$,
where $L_{\rm iso}$ is the isotropic luminosity at the peak
of the prompt light curve, but calculated using the
{\it time averaged} spectrum (i.e. \ep\ and spectral indices),
and not the spectral properties at the peak flux. 

This correlation appeared to be tighter (but with similar slope) than
the \ama\ correlation, as originally found by Amati et al. (2002).
Since then this correlation
has been updated only once (Ghirlanda et al., 2005b). 

It is interesting to test if the same conclusions that can be drawn
for the \ama\ correlation (i.e. the presence of selection effects and
of outliers) can now be extended to the \yon\ correlation.  To this
aim we have considered the $z$GRB sample (see Tab. \ref{tabz} in the
Appendix) and we have calculated for all these bursts their isotropic
equivalent luminosity $L_{\rm iso}$.  This is computed by integrating
the time averaged spectrum after renormalizing it with the peak flux.
Note that, strictly speaking, this luminosity does not correspond to
the peak luminosity (see Ghirlanda et al. 2005b), since it adopts the
time averaged \epo.

In Tab. \ref{tabz} we report the sample of 83 GRBs with their peak flux, the
energy range where it is computed, the references and $L_{\rm iso}$. 
To calculate $L_{\rm iso}$ we adopted the same method used to 
compute $E_{\rm iso}$ (see Ghirlanda et al. 2007 for more details).

In Fig. \ref{yone} we show the \yon\ and the \ama\ correlations defined with
the 83 GRBs of Tab. \ref{tabz}.  The no--\sw\ bursts (empty symbols) and the
\sw\ bursts (filled squares) are shown. In both cases, the correlations are
highly significant (rank correlation coefficient are respectively 0.83 with a
chance probability $6.2\times 10^{-22}$ and 0.84 with a chance probability
$9.8\times 10^{-23}$).  The solid lines show the best fit with the least
square method (without accounting for the measurement errors): we obtain
\ep$\propto L_{\rm iso}^{0.40\pm0.03}$ and \ep$\propto E^{0.48\pm0.03}_{\rm
  iso}$. The fits of the no--\sw\ burst sample (dotted line) and of the \sw\ 
burst sample (dashed line) are also shown. The results of these fits performed
considering different samples are shown in Tab. \ref{tab1}.

Our sample of 83 $z$GRBs confirms the finding of Yonetoku et al. (2004),
even if we obtain a flatter slope. Fitting the
scatter distribution of the \yon\ correlation with a Gaussian we derive
$\sigma=0.28$. Comparing it with the corresponding scatter of the \ama\ 
correlation ($\sigma=0.23$) we find that, contrary to what initially
found by Yonetoku et al. (2004), the scatter of this correlation is slightly
larger.
 
We can investigate if this correlation is affected by any of the
selections effects that have been studied in G08 for the \ama\
correlation. In particular we show in Fig. \ref{outliers_yone} the
observer frame \yonf\ correlation where $P$ is the bolometric peak
flux. Note that also in this plane the $z$GRBs define a strong
correlation (dotted line -- with slope 0.39) and that the GRB samples
without $z$ considered in this work are consistent with this
correlation (differently to what happens in the \amaf\ plane).
The dashed line represents the best fit obtained considering only \ba\
bursts.  They define a flatter correlation (slope 0.28) with respect
to the $z$GRB sample.  Note that this happens also in the \amaf\
correlation and it is likely due to the difficulty of the \ba\
instrument to see very low \epo\ at low fluence/peak flux.

The peak flux $P$ is the quantity on which the trigger condition (for most
instruments) is determined. 
We plot in Fig. \ref{outliers_yone} the trigger
limiting curves (from G08) as a function of \epo. We note that for \ba\ 
the TT curve is separated from the distribution of the
corresponding bursts (open and filled circles).  This is because the dominant
selection effect acting on our \ba\ complete sample is the ST  (see
Fig. \ref{solobatse}). In other words, the bursts that can be displayed in the
\yonf\ plane are not all the bursts that can be detected by a given
instrument, but only those with a sufficient number of photons to make
possible the determination of \epo.

The \he\ bursts, instead (triangles), are very near to their TT. 
For this instrument we are not able to determine the ST
curves, but it is likely that the dominant selection effect acting on \he\ 
bursts is the need to trigger them.

For  \sw\ bursts we have an intermediate case:
their TT curve is not truncating their distribution,
even if they lie closer (than \ba\ bursts) to it.

Also for the \yon\ correlation we can test if there are outliers.
Ghirlanda et al. (2005b) tested this through a sample of 442 GRBs with
redshifts derived by the lag--luminosity relation. They did not find
evidence for outliers. In this work we test the \yon\ correlation with
the same method described above for the \ama\ correlation.  In
Fig. \ref{outliers_yone} we show the ``region of outliers'' for the
\yon\ correlation. Only one burst (of the K06 sample) is inconsistent
with this correlation at more than 3$\sigma$.

\section{Discussion and conclusions}

\begin{figure}
\hskip -0.5 true cm
\psfig{figure=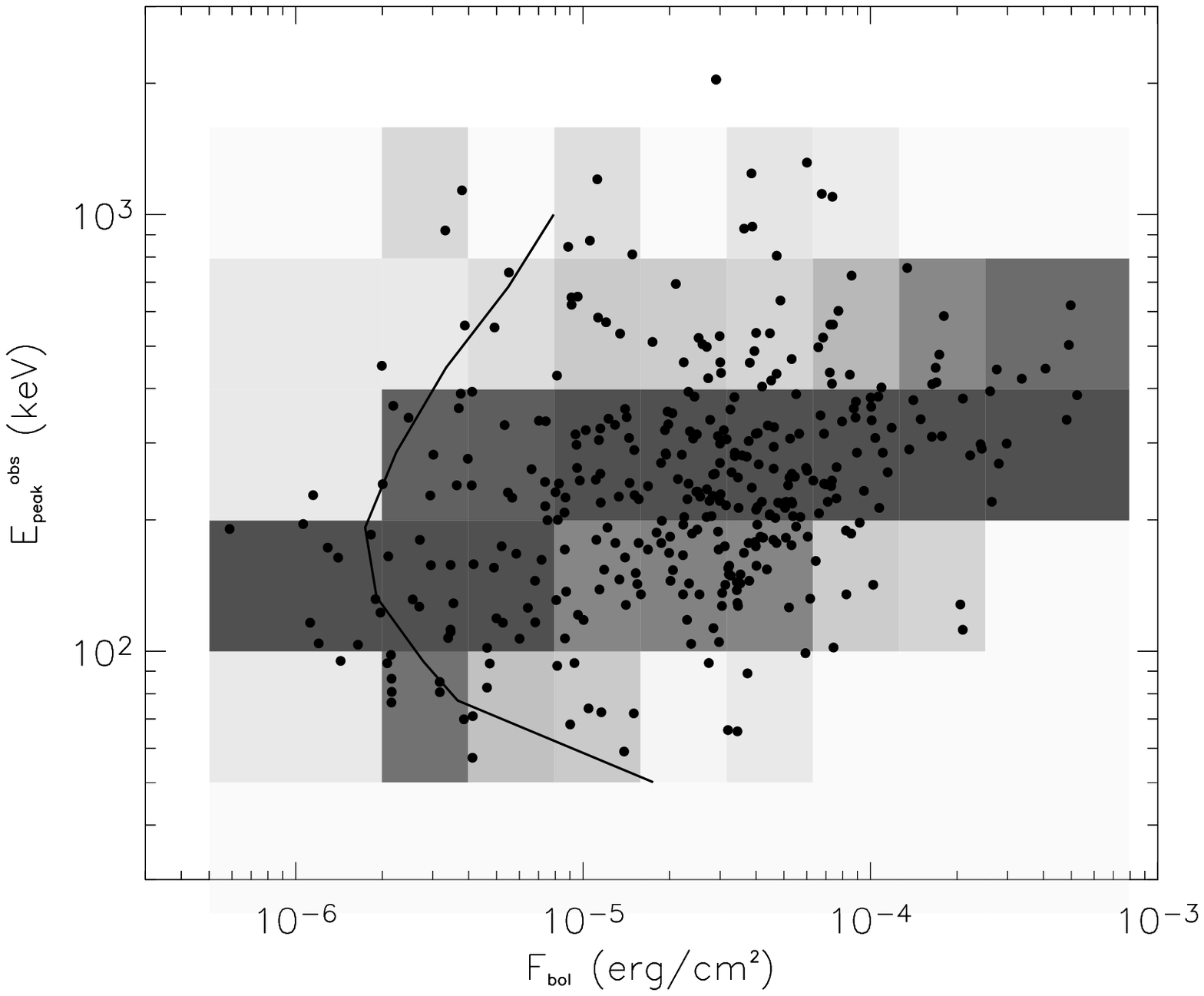,width=9cm,height=7.5cm}
\caption{
Graphic illustration of the \amaf\ correlation.
We consider different fluence--bins and in each of those we count
the total number $N_f$ of objects. 
Then we divide this fluence--bin into \epo--bins,
counting the number of objects in each small area, dividing it
by $N_f$.
Each small area is then characterized by a number $n_{i,f}$ (between 0 and 1),
which corresponds to a different level of grey.
In this way the increasing number of bursts for decreasing fluence 
(the ``Log$N$--Log$S$" effect) is accounted for, and it does not influence 
$n_{i,f}$.  
Note that the data points do not fill the entire accessible region of
the plane (to the right of the ST curve shown in Fig. \ref{solobatse}),
but concentrate along a stripe.
}
\label{fettine}
\end{figure}

To study the role of possible instrumental selection effects on the
Amati relation we have focused our attention on the observational
\amaf\ plane. 
Here we can compare the distribution of different samples of
GRB (for example, $z$GRBs and GRB with unknown redshift).
To this aim we adopt the analysis performed by G08, referring to two
different instrumental biases: the trigger threshold (TT, the minimum
fluence derived considering the minimum flux required to trigger a burst) and
the spectral analysis threshold (ST, the
minimum fluence needed to constrain the GRB spectral properties). 
These curves depends on \epo\ and
define what part of the observational plane is accessible.

First we updated the sample of bursts with redshift, adding 7 
new recent GRBs, for a total of 83 objects.
These GRBs define a \ep$\propto$\eiso$^{0.48\pm 0.03}$ 
correlation in the rest frame, very similar to that obtained with 
previous (and smaller) samples.
In the observer plane, they define a slightly flatter correlation
(\epo$\propto F^{0.40\pm0.05}$).
The scatter of these two correlations is the same (see Tab. \ref{tab1}).
As G08 pointed out, the \ba\ ST curve is not biasing
the distribution of \ba\ bursts with redshift in the observer plane,
while the \sw\ ST could, in the sense that the
distribution of \sw\ bursts (with redshift) is truncated
by the \sw\ ST curve.
Then why the \ba\ bursts (with redshift)
are not truncated by their corresponding ST?
Is it because of a real, intrinsic correlation or 
is it due to another, hidden, selection effect?
One way to answer this crucial question is to 
analyze all GRBs with \epo, even without redshift.
The \ba\ sample of GRBs is the best suited for this aim
because: i) it contains a large number of bursts;
ii) large sample of bright GRBs have been already analyzed,
and iii) for \ba\ we already know the ST curve.
Then we pushed the spectral analysis to the limit,
deriving the spectral parameters for a representative sample of 
100 \ba\ GRBs with  a (bolometric) fluence between $10^{-6}$ [erg cm$^{-2}$]
(corresponding to the ST limit) and $2\times 10^{-5}$ [erg cm$^{-2}$]
(the limiting fluence of K06).
These 100 GRBs represent a large population of 1000 GRBs,
in the same fluence range.
Combining our and the K06 samples we have a homogeneous and
complete sample, best suited to study how \ba\ GRBs populate
the \amaf\ plane.
Using this complete, fluence limited, sample we find:

\begin{itemize}
\item
GRBs without redshifts, in this plane, are not spread in the 
region free from instrumental selection effects, but define a correlation 
with a flat slope ($\sim 0.16$) and a scatter
larger for smaller fluences (after accounting for the errors increasing 
for smaller fluences). 
Fig. \ref{fettine} is a graphic illustration of this:
different grey levels corresponds to different density of points
in the \amaf\ plane, once we take out the effect of the overall
increase in density going to smaller fluences
(for the Log$N$--Log$S$ slope). 
The way we do this is the following:
we consider different fluence--bins and in each of those we count
the total number $N_f$ of objects. 
Then we divide this fluence--bin into \epo--bins,
counting the number of objects in each small area, dividing it
by $N_f$.
Each small area is then characterized by a number $n_{i,f}$ (between 0 and 1),
which corresponds to a different level of grey.
Note that the data points do not fill the entire accessible region of
the plane but concentrate along a stripe.  Note that the shape of this
  concentration of points is not determined by the ST curve, reported
  in Fig. \ref{fettine} for a typical burst lasting 20 s. The only
effect of the ST curve to the found correlation is to cut it at the
smallest fluences and \epo.  The very flat slope could be due to the
difficulty of having \ba\ GRBs with \epo\ smaller than $\sim 50$ keV,
whose existence is demonstrated by other instruments.  However, the
paucity of the derived upper limits on \epo\ suggests that this effect
is marginal.

\item
Formally, the scatter is not greater than the scatter
of the \ama\ correlation (both have $\sigma=0.23$ once the
contribution to the scatter of the measurement errors are
taken into account).
Despite that, the entire \ba\ sample and the $z$GRB population
define two \amaf\ correlations which have significantly different
slopes.
If their redshift distribution is similar, then they will define
two different correlations also in the \ama\ plane: considering then
the two samples together, we will define a correlation with intermediate 
slope and a scatter larger than the individual one.

\item
If the above point holds (i.e. if the redshift distributions
of GRBs of unknown redshifts is the same of the $z$GRBs)
then we can conclude that there exists an \ama\  correlation,
not determined by selection effects, even if its slope and scatter
may be different from what we know now.
We should emphasize that by the term "correlation" 
we mean that GRBs will occupy a ``stripe" in the
\ama\ plane with a relatively large scatter
(fitting it with $\chi^2$ method one would obtain a
very large reduced $\chi^2_r$).
In other words, it is very likely that there is another (third)
variable responsible for the scatter.
In fact one finds a tighter correlation considering,
as a third variable, the jet break time (Ghirlanda, Ghisellini \& Lazzati 2004;
Laing \& Zhang 2005) or the time of enhanced prompt emission (Firmani et al. 2005).
Another cause for a large scatter is the viewing angle,
if a significant number of bursts are seen slightly off--axis.

\item
In the \ba\ sample there are a few bursts with small or intermediate fluences 
but large \epo, not present in the $z$GRB sample.
Among them there are some surely outliers of the \ama\ correlation 
(as defined by the $z$GRB sample), i.e. bursts that lie at more than 
3$\sigma$  from it, for any redshift.
The number of these sure outliers is however very small,
amounting to the 6 per cent of the entire population.

\item
We have also investigated the \yon\ correlation,
and its counterpart (\yonf) in the observer plane.
First, we partly confirm the original findings of Yonetoku et al. (2005,
see also the update in Ghirlanda et al. 2005b):
with the $z$GRB sample we find a strong \yon\ correlation, whose
slope is flatter than originally found ($s=0.40$ instead of 0.5)
and whose scatter is greater than the scatter of the \ama\ correlation.

\item
In the observer plane, instead, the \yonf\ correlation of our complete 
sample of \ba\ bursts is tighter than the the \amaf\ correlation
($\sigma=0.2$ instead of $\sigma=0.23$).
Its slope is $s=0.24$, flatter than the \yon\ correlation ($s=0.39$), 
but however steeper than the \amaf\ slope ($s=0.15$).
There is only one sure outlier.

\item
Selection effects are in this case determined by the TT curves.
These effects are present,
being responsible for the cutting at low peak fluxes,
but they do not influence the slope and scatter for peak
fluxes larger than the what defined by the TT curves.

\item
Considering the $z$GRB sample we have that the
\ama\ correlation is tighter than the \yon\ one.
Considering our complete \ba\ sample and moving to 
the observer plane, we have just the opposite:
the \yonf\ correlation is tighter than the \amaf\ one.

\item 
It is then conceivable that the \yon\ correlation,
once a large number of burst with redshift will be available,
will be stronger than the \ama\ one.

\end{itemize}

The general conclusion we can draw from our study is that,
although selection effects are present, they do not determine the
spectral--energy and spectral--luminosity correlations.
These could be characterized by a slope and scatter
different from what we have determined now using heterogeneous
bursts samples with measured redshift, but we found that \ep\ is
indeed correlated with the burst energetics or peak luminosity.
Therefore it is worth to investigate the physical reason for this
relation.

\section*{Acknowledgements} 
We thank M. Nardini for stimulating discussions. We thank partial
funding by a 2008 PRIN--INAF grant and ASI I/088/06/0 for funding.

\appendix
\section{Tables}
\vskip 5.0 true cm
\setcounter{table}{0}
\begin{table*}
\begin{tabular}{llccllrl}
\hline
GRB     &  $\alpha$               & $\beta$                 &  \epo\                  &  F          &   P      &  T     &  GCN\\
        &                         &                         &  keV                    &erg/cm$^{2}$ &erg/s/cm$^2$&s     &     \\
\hline
050326  & -0.74$\pm0.09$          & -2.49$\pm0.16$          &   201$\pm24$            &    3.6e-5   &  6.8e-6  &    38  & 3152\\
050713A & -1.12$\pm0.08$          &                         &   312$\pm50$            &    1.3e-5   &  1.8e-6  &    16  & 3619\\
050717  & -1.12$^{+0.17}_{-0.13}$ &                         &  1890$^{+1600}_{-760}$  &    6.3e-5   &  1.2e-5  &    50  & 3640\\ 
051008  & -0.98$^{+0.09}_{-0.08}$ &                         &   865$^{+178}_{-136}$   &    5.5e-5   &  9.8e-6  &   280  & 4078\\
051028  & -0.73$^{+0.26}_{-0.22}$ &                         &   298$^{+73}_{-50}$     &    7.0e-6   &  1.3e-6  &    12  & 4183\\
060105  & -0.83$\pm0.04$          &                         &   424$^{+25}_{-22}$     &    8.1e-5   &  6.7e-6  &    60  & 4439\\
060213  & -0.83$^{+0.05}_{-0.04}$ &                         &  1061$^{+83}_{-43}$     &    1.6e-4   &  7.1e-5  &    60  & 4763\\
060510A & -1.66$^{+0.08}_{-0.07}$ &                         &   184$^{+36}_{-24}$     &    3.6e-5   &  6.1e-6  &    25  & 5113\\
060901  & -0.77$^{+0.26}_{-0.23}$ & -2.31$^{+0.18}_{-0.36}$ &   191$^{+40}_{-30}$     &    1.9e-5   &  3.8e-6  &     8  & 5498\\
060904A & -1.00$^{+0.23}_{-0.17}$ & -2.57$^{+0.37}_{-1.00}$ &   163$\pm31$            &    1.9e-5   &  1.6e-6  &    80  & 5518\\
060928  & -1.28$\pm0.02$          & -2.27$^{+0.14}_{-0.21}$ &   705$^{+74}_{-68}$     &    3.8e-4   &  3.1e-5  &        & 5689\\
061021  & -1.22$^{+0.14}_{-0.12}$ &                         &   777$^{+549}_{-237}$   &    1.1e-5   &  4.3e-6  &        & 5748\\
061122  & -1.03$^{+0.07}_{-0.06}$ &                         &   160$^{+8}_{-7}$       &    2.6e-5   &  9.9e-6  &    10  & 5841\\
061222A & -0.94$^{+0.14}_{-0.13}$ & -2.41$^{+0.28}_{-1.21}$ &   283$^{+59}_{-42}$     &    3.3e-5   &  7.0e-6  &   100  & 5984\\
070220  & -1.21$^{+0.29}_{-0.19}$ & -2.02$^{+0.27}_{-0.44}$ &   299$^{+204}_{-130}$   &    4.4e-5   &  3.7e-6  &   130  & 6124\\
070328  & -1.09$\pm0.08$          &                         &   688$^{+173}_{-119}$   &    6.1e-5   &  4.7e-6  &    45  & 6230\\
070402  & -0.92$^{+0.12}_{-0.11}$ &                         &   325$^{+52}_{-40}$     &    1.2e-5   &  3.3e-6  &    12  & 6243\\
070521  & -0.93$\pm0.12$          &                         &   222$^{+27}_{-21}$     &    1.9e-5   &  4.4e-6  &    55  & 6459\\
070626  & -1.45$^{+0.04}_{-0.03}$ & -2.28$^{+0.08}_{-0.12}$ &   226$^{+19}_{-17}$     &    4.9e-4   &  2.4e-5  &        & 6599\\
070724B & -1.15$\pm0.13$          &                         &    82$\pm5$             &    2.3e-5   &  2.8e-6  &    50  & 6671\\
070821  & -1.30$\pm0.04$          &                         &   268$^{+19}_{-17}$     &    1.2e-4   &  1.5e-5  &   215  & 6766\\
070824  & -1.05$\pm0.08$          &                         &   253$^{+22}_{-19}$     &    3.1e-5   &  3.0e-5  &    12  & 6768\\
070917  & -1.36$^{+0.25}_{-0.21}$ &                         &   211$^{+95}_{-48}$     &    5.9e-6   &  5.4e-6  &     6  & 6798\\
071006  & -0.84$^{+0.26}_{-0.22}$ &                         &   334$^{+95}_{-61}$     &    2.2e-5   &  2.7e-6  &    60  & 6867\\
071125  & -0.62$\pm0.05$          & -3.10$^{+0.25}_{-0.41}$ &   299$\pm13$            &    7.8e-5   &  2.4e-5  &        & 7137\\
080122  & -1.21$^{+0.12}_{-0.11}$ & -2.36$^{+0.24}_{-0.68}$ &   277$^{+43}_{-33}$     &    9.5e-5   &  7.1e-6  &   150  & 7219\\
080204  & -1.35$^{+0.06}_{-0.09}$ &                         &  1279$^{+469}_{-382}$   &    2.8e-5   &  1.0e-5  &        & 7263\\
080211  & -0.85$\pm0.06$          &                         &   356$^{+25}_{-22}$     &    4.8e-5   &  6.9e-6  &        & 7309\\
080328  & -1.13$^{+0.17}_{-0.20}$ &                         &   289$^{+93}_{-57}$     &    2.5e-5   &  2.8e-6  &    90  & 7548\\

\hline
\end{tabular}
\caption{Spectral and temporal properties of 29 \ko\ bursts without
  known redshift. The listed fluence and peak flux are estimated in
  the range 1--10$^4$ keV. It is also reported (when available) the
  estimated burst duration T. The references are: Golenetskii et al.,
  2005a, 2005b, 2005c, 2005d, 2005e; Golenetskii et al., 2006a, 2006b,
  2006c, 2006d, 2006e, 2006f, 2006g, 2006h, 2006i; Golenetskii et al.,
  2007a, 2007b, 2007c, 2007d, 2007e, 2007f, 2007g, 2007h, 2007i,
  2007j, 2007k; Golenetskii et al., 2008a, 2008b, 2008c, 2008d. The
  GCN circulars number is reported in the last column. }
\label{tabkonus}
\end{table*}

\begin{table*}
\begin{tabular}{rlllllrllll}
\hline
GRB         &$\alpha$       &$\beta$         & $E^{\rm {obs}}_{\rm peak}$ & Fluence   &   &  GRB &$\alpha$    &$\beta$  & $E^{\rm{obs}}_{\rm peak}$ & Fluence \\
            &               &                &  keV                       & erg/cm$^2$&   &      &            &         &  keV                      & erg/cm$^2$\\
\hline
469      &-1.16$\pm$0.055   &               &581   $\pm$ 95     &(1.1$\pm$0.2)e--5&     &  5419     &-1.52$\pm$ 0.12 &               &106  $\pm$  29    &(6.0$\pm$1.3)e--6  \\
658      &-1.71$\pm$0.28    &-2.30          &70    $\pm$ 56     &(3.8$\pm$1.9)e--6&     &  5428     &-1.20           &               &103  $\pm$  68    &(1.6$\pm$1.2)e--6  \\
803      &-0.71$\pm$0.18    &               &241   $\pm$ 66     &(2.0$\pm$0.7)e--6&     &  5454     &-1.00           &-2.30          &71   $\pm$  8.2   &(4.1$\pm$0.7)e--6  \\
829      &-0.07$\pm$0.33    &-3.26$\pm$0.69 &127   $\pm$ 29     &(1.4$\pm$0.9)e--5&     &  5464     &-0.70$\pm$0.17  &               &329  $\pm$  91    &(5.3$\pm$1.9)e--6  \\
938      &-0.96$\pm$0.36    &               &110   $\pm$ 46     &(3.5$\pm$2.3)e--6&     &  5466     &-1.00           &-2.59$\pm$0.11 &$>$50$\pm$        &(2.6$\pm$1.1)e--6  \\
1025     &-0.17$\pm$0.77    &-2.22$\pm$0.21 &117   $\pm$ 74     &(1.0$\pm$0.4)e--5&     &  5467     &-0.23$\pm$0.31  &-2.54$\pm$2.53 &450  $\pm$  229   &(2.0$\pm$1.8)e--6  \\
1406     &-0.19$\pm$0.93    &               &116   $\pm$ 94     &(5.3$\pm$4.9)e--6&     &  5484     &-1.20           &               &336  $\pm$  142   &(7.4$\pm$2.4)e--6  \\
1425     &-1.54$\pm$0.03    &               &153   $\pm$ 15     &(1.2$\pm$0.1)e--5&     &  5493     &-1.20           &               &157  $\pm$  63    &(3.5$\pm$1.3)e--6  \\
1447     &-0.46$\pm$0.08    &-3.02$\pm$0.44 &247   $\pm$ 24     &(1.1$\pm$0.2)e--5&     &  5518     &-1.04$\pm$0.07  &               &155  $\pm$  16    &(4.9$\pm$0.6)e--6  \\
1559     &-0.35$\pm$0.31    &-2.05$\pm$0.27 &224   $\pm$ 84     &(5.7$\pm$3.3)e--6&     &  5538     &0.21 $\pm$0.32  &               &227  $\pm$  58    &(1.1$\pm$0.6)e--6  \\ 
1586     &0.74 $\pm$0.68    &-3.24$\pm$0.49 &94    $\pm$ 28     &(4.7$\pm$3.2)e--6&     &  5541     &-0.99$\pm$0.37  &               &131  $\pm$  63    &(2.5$\pm$2.0)e--6  \\
1660     &-0.85$\pm$0.31    &               &101   $\pm$ 31     &(4.6$\pm$2.6)e--6&     &  5593     &-0.73$\pm$ 0.09 &-3.17$\pm$1.47 &199  $\pm$  29    &(7.5$\pm$2.7)e--6  \\
1667     &                  &-4.75$\pm$0.25 &$<$30              &(1.1$\pm$0.1)e--6&     &  5721     &-1.10$\pm$0.35  &               &844  $\pm$  756   &(8.9$\pm$6.0)e--6  \\
1683     &-1.17$\pm$0.04    &               &337   $\pm$ 36     &(7.0$\pm$0.5)e--6&     &  5725     &-1.00           &-2.30          &304  $\pm$  22    &(1.1$\pm$0.1)e--5  \\
1717     &-0.94$\pm$0.11    &-2.58$\pm$0.19 &167   $\pm$ 26     &(5.9$\pm$1.2)e--6&     &  5729     &-1.33$\pm$0.76  &-2.08$\pm$0.03 &65   $\pm$  6.7   &(3.4$\pm$1.2)e--5  \\
1956     &-1.20$\pm$0.13    &-2.44$\pm$0.19 &125   $\pm$ 31     &(6.4$\pm$1.8)e--6&     &  6083     &-1.12$\pm$ 0.26 &-2.78$\pm$0.60 &98   $\pm$  36    &(2.1$\pm$1.4)e--6  \\
2093     &                  &               &$<$30              &(3.4$\pm$0.7)e--6&     &  6090     &-0.94$\pm$ 0.09 &-3.38$\pm$1.33 &158  $\pm$  22    &(4.2$\pm$1.2)e--6  \\
2123     &-1.42$\pm$0.07    &               &94    $\pm$ 11     &(9.3$\pm$1.2)e--6&     &  6098     &-0.62$\pm$0.16  &               &122  $\pm$  19    &(2.0$\pm$0.6)e--6  \\
2315     &-0.99$\pm$0.37    &               &242   $\pm$ 156    &(8.2$\pm$6.1)e--6&     &  6104     &                &-4.00$\pm$1.45 &$<$30             &(1.1$\pm$0.4)e--6  \\
2430     &-1.20             &               &275   $\pm$ 211    &(4.7$\pm$2.4)e--6&     &  6216     &-1.20           &               &95   $\pm$  16    &(1.4$\pm$0.3)e--6  \\
2432     &-1.46$\pm$0.08    &-2.30          &$>$107             &(4.1$\pm$0.6)e--6&     &  6251     &-1.16$\pm$0.10  &               &557  $\pm$  205   &(3.9$\pm$1.4)e--6  \\
2443     &-0.78$\pm$0.45    &-2.08$\pm$0.39 &244   $\pm$ 183    &(7.4$\pm$6.5)e--6&     &  6303     &-0.98$\pm$0.13  &-2.40$\pm$0.39 &227  $\pm$  60    &(1.5$\pm$0.5)e--5  \\
2447     &                  &               &$<$30              &(3.4$\pm$1.3)e--6&     &  6399     &-1.20           &               &81   $\pm$  35    &(2.2$\pm$1.1)e--6  \\
2458     &                  &-2.60          &$<$30              &(1.5$\pm$0.6)e--6&     &  6405     &-1.20           &               &157  $\pm$  80    &(2.9$\pm$1.5)e--6  \\
2476     &-1.31$\pm$0.53    &-3.33$\pm$2.78 &59    $\pm$ 35     &(1.4$\pm$1.0)e--5&     &  6450     &                &               &$<$30             &(1.6$\pm$1.1)e--6  \\
2640     &-1.30             &-2.30          &131   $\pm$ 29     &(1.9$\pm$0.4)e--6&     &  6521     &-1.21$\pm$0.38  &-2.94$\pm$3.27 &116  $\pm$  77    &(6.8$\pm$5.5)e--6  \\
2736     &-1.20             &               &112   $\pm$ 55     &(3.4$\pm$1.8)e--6&     &  6523     &                &               &$<$30             &(2.8$\pm$0.6)e--6  \\
2864     &-0.81$\pm$0.22    &-2.06$\pm$0.24 &239   $\pm$ 99     &(4.1$\pm$1.9)e--6&     &  6550     &-1.20           &               &282  $\pm$  156   &(3.0$\pm$1.1)e--6  \\
3001     &-1.05$\pm$0.12    &-2.11$\pm$0.16 &219   $\pm$ 62     &(1.2$\pm$0.3)e--5&     &  6611     &-1.20           &               &116  $\pm$  93    &(1.1$\pm$0.9)e--6  \\
3032     &-0.34$\pm$0.24    &-2.86$\pm$0.43 &126   $\pm$ 26     &(2.7$\pm$1.6)e--6&     &  6621     &-1.44$\pm$0.26  &               &74   $\pm$  28    &(1.0$\pm$0.5)e--5  \\
3056     &-1.64$\pm$0.09    &-2.57$\pm$0.75 &135   $\pm$ 52     &(1.6$\pm$0.4)e--5&     &  6672     &-1.78$\pm$0.08  &-2.46$\pm$0.16 &72   $\pm$  21    &(1.5$\pm$0.3)e--5  \\
3075     &-1.46$\pm$0.19    &-2.42$\pm$0.19 &93    $\pm$ 38     &(8.1$\pm$3.3)e--6&     &  6764     &-1.45$\pm$0.09  &               &343  $\pm$  141   &(1.4$\pm$0.4)e--5  \\
3091     &-1.20             &               &104   $\pm$ 19     &(1.2$\pm$0.3)e--6&     &  6824     &0.35 $\pm$0.70  &-2.06$\pm$0.89 &342  $\pm$  220   &(2.5$\pm$2.1)e--6  \\
3093     &-1.20             &               &261   $\pm$ 225    &(6.6$\pm$4.3)e--6&     &  7290     &-0.33$\pm$0.25  &               &1136 $\pm$  582   &(3.8$\pm$2.0)e--6  \\
3101     &-1.59$\pm$0.22    &-2.01$\pm$0.09 &138   $\pm$ 21     &(1.1$\pm$0.3)e--5&     &  7319     &-0.23$\pm$0.5   &-2.30          &121  $\pm$  81    &(9.6$\pm$8.0)e--6  \\
3177     &-1.09$\pm$0.18    &-2.55$\pm$0.52 &164   $\pm$ 55     &(2.1$\pm$0.8)e--6&     &  7374     &-0.81$\pm$0.42  &-2.26$\pm$0.84 &207  $\pm$  142   &(8.6$\pm$7.5)e--6  \\
3217     &-1.60             &-2.20          &231   $\pm$ 151    &(8.0$\pm$2.1)e--6&     &  7387     &0.51 $\pm$1.58  &-2.34$\pm$0.21 &83   $\pm$  69    &(4.6$\pm$4.2)e--6  \\
3220     &-0.19$\pm$0.60    &               &179   $\pm$ 100    &(2.7$\pm$2.0)e--6&     &  7504     &-1.34$\pm$0.22  &-2.61$\pm$0.50 &107  $\pm$  43    &(3.4$\pm$1.8)e--6  \\
3276     &-1.00             &-2.30          &190   $\pm$ 53     &(5.9$\pm$1.8)e--7&     &  7552     &                &               &$<$30             &(1.5$\pm$0.7)e--6  \\
3319     &-1.02$\pm$0.37    &               &195   $\pm$ 134    &(1.1$\pm$0.9)e--6&     &  7597     &-1.20           &               &163  $\pm$  141   &(1.4$\pm$1.0)e--6  \\
3516     &-1.23$\pm$0.17    &-2.28$\pm$0.66 &313   $\pm$ 180    &(9.4$\pm$3.8)e--6&     &  7638     &-0.66$\pm$0.52  &-2.67$\pm$0.09 &57   $\pm$  22    &(4.1$\pm$3.2)e--6  \\
3552     &                  &-3.03$\pm$0.14 &$<$30              &(2.3$\pm$0.9)e--6&     &  7677     &0.01$\pm$0.90   &-2.81$\pm$0.53 &86   $\pm$  47    &(2.2$\pm$2.0)e--6  \\
3569     &-1.47$\pm$0.14    &               &227   $\pm$ 114    &(2.9$\pm$0.9)e--6&     &  7684     &-0.58$\pm$0.20  &-2.10          &646  $\pm$  257   &(9.1$\pm$4.5)e--6  \\
3869     &-1.20             &               &172   $\pm$ 53     &(1.3$\pm$0.4)e--6&     &  7750     &-1.20           &               &76   $\pm$  14    &(2.2$\pm$0.5)e--6  \\
3875     &                  &-3.13$\pm$0.69 &$<$30              &(1.1$\pm$0.2)e--6&     &  7769     &-1.50$\pm$0.17  &               &72   $\pm$  26    &(1.2$\pm$0.5)e--5  \\
3893     &-0.65$\pm$0.04    &               &153   $\pm$ 6.3    &(2.0$\pm$0.1)e--5&     &  7781     &-0.66$\pm$0.77  &               &94   $\pm$  61    &(2.1$\pm$1.9)e--6  \\
4048     &-0.63$\pm$0.08    &-2.87$\pm$0.45 &323   $\pm$ 42     &(1.1$\pm$0.4)e--5&     &  7838     &-0.95$\pm$0.35  &               &239  $\pm$  153   &(3.6$\pm$2.6)e--6  \\
4146     &-0.55$\pm$0.57    &-3.12$\pm$0.51 &85    $\pm$ 34     &(3.2$\pm$1.5)e--6&     &  7845     &-0.09$\pm$0.42  &               &551  $\pm$  291   &(4.9$\pm$3.5)e--6  \\
4216     &-1.20             &               &128   $\pm$ 64     &(3.5$\pm$1.7)e--6&     &  7989     &-1.52$\pm$0.11  &-2.30          &360  $\pm$  216   &(3.7$\pm$0.9)e--6  \\
5417     &-0.71$\pm$0.19    &-2.01$\pm$0.06 &142   $\pm$ 41     &(1.5$\pm$0.6)e--5&     &  7998     &-1.26$\pm$ 0.31 &               &81   $\pm$  37    &(3.2$\pm$2.0)e--6  \\

\hline
\end{tabular}
\caption{Sample of \ba\ bursts analyzed in this work. 
The trigger number and the spectral parameters of the fit of the time 
integrated spectrum are reported. In the last column we report the bolometric
fluence obtained by integrating the best fit spectrum. For those bursts whose
spectrum allows only to set a lower/upper limit on \epo\ we report
the \ba\ catalogue fluence (i.e. $>$25 keV). When is not
possible to constrain the value of $\alpha$ we performed the
spectral fit by fixing $\alpha$ to an appropriate value. These values are
reported in table without errors.}
\label{tabverdi}
\end{table*}

\begin{table*}
\begin{tabular}{lllllllllll}
\hline 
GRB        &$z$    &$\alpha$        &$\beta$         &Peak Flux$^a$       &Range     &$L_{\rm iso}$  & $E_{\rm peak}$    & Ref \\ 
           &       &                &                &                    &keV       &erg/s            & keV             &     \\
\hline
970228   &0.695  &-1.54 [ 0.08 ]&  -2.5   [ 0.4  ]&     3.7e-6     [   0.8e-6   ] &    40-700      &   9.1e51   [ 2.18e51   ]&    195   [ 64    ]   &    1  \\
970508   &0.835  &-1.71 [ 0.1  ]&  -2.2   [ 0.25 ]&     7.4e-7     [   0.7e-7   ] &    50-300      &   9.4e51   [ 1.25e51   ]&    145   [ 43    ]   &    3  \\
970828   &0.958  &-0.7  [ 0.08 ]&  -2.1   [ 0.4  ]&     5.9e-6     [   0.3e-6   ] &    30-1.e4     &   2.51e52  [ 7.7e51    ]&    583   [ 117   ]   &    1  \\
971214   &3.42   &-0.76 [ 0.1  ]&  -2.7   [ 1.1  ]&     6.8e-7     [   0.7e-7   ] &    40-700      &   7.21e52  [ 1.33e52   ]&    685   [ 133   ]   &    1  \\
980326   &1.0    &-1.23 [ 0.21 ]&  -2.48  [ 0.31 ]&     2.45e-7    [   0.15e-7  ] &    40-700      &   3.47e51  [ 1.e51     ]&    71    [ 36    ]   &    4  \\
980613   &1.096  &-1.43 [ 0.24 ]&  -2.7   [ 0.6  ]&     1.6e-7     [   0.4e-7   ] &    40-700      &   2.e51    [ 6.7e50    ]&    194   [ 89    ]   &    4  \\
980703   &0.966  &-1.31 [ 0.14 ]&  -2.39  [ 0.26 ]&     1.6e-6     [   0.2e-6   ] &    50-300      &   2.09e52  [ 4.86e51   ]&    499   [ 100   ]   &    1  \\
990123   &1.600  &-0.89 [ 0.08 ]&  -2.45  [ 0.97 ]&     1.7e-5     [   0.5e-5   ] &    40-700      &   3.53e53  [ 1.23e53   ]&    2031  [ 161   ]   &    1  \\
990506   &1.307  &-1.37 [ 0.15 ]&  -2.15  [ 0.38 ]&     18.6       [   0.1      ] &    50-300      &   4.18e52  [ 1.33e52   ]&    653   [ 130   ]   &    1  \\
990510   &1.619  &-1.23 [ 0.05 ]&  -2.7   [ 0.4  ]&     2.5e-6     [   0.2e-6   ] &    40-700      &   6.12e52  [ 1.07e52   ]&    423   [ 42    ]   &    1  \\
990705   &0.843  &-1.05 [ 0.21 ]&  -2.2   [ 0.1  ]&     3.7e-6     [   0.1e-6   ] &    40-700      &   1.65e52  [ 2.77e51   ]&    348   [ 28    ]   &    1  \\
990712   &0.433  &-1.88 [ 0.07 ]&  -2.48  [ 0.56 ]&     4.1        [   0.3      ] &    40-700      &   7.46e50  [ 1.91e50   ]&    93    [ 15    ]   &    1  \\
991208   &0.706  &              &                 &     1.85e-5    [   0.06e-5  ] &    20-1.e4     &   4.32e52  [  0.38e52  ]&    313   [ 31    ]   &    2  \\
991216   &1.02   &-1.23 [ 0.13 ]&  -2.18  [ 0.39 ]&     67.5       [   0.2      ] &    50-300      &   1.13e53  [  3.75e52  ]&    642   [ 129   ]   &    1  \\
000131   &4.50   &-0.69 [ 0.08 ]&  -2.07  [ 0.37 ]&     7.89       [   0.08     ] &    50-300      &   1.41e53  [  5.59e52  ]&    714   [ 142   ]   &    1  \\
000210   &0.846  &              &                 &     2.42e-5    [   0.15e-5  ] &    20-1.e4     &   8.78e52  [  1.1e52   ]&    753   [ 26    ]   &    2  \\
000418   &1.12   &              &                 &     2.8e-6     [   0.4e-6   ] &    20-1.e4     &   2.e51    [  4.8e50   ]&    284   [ 21    ]   &    2  \\
000911   &1.06   &-1.11 [ 0.12 ]&  -2.32  [ 0.41 ]&     2.0e-5     [   0.2e-5   ] &    15-8000     &   1.65e53  [  2.89e52  ]&    1856  [ 371.  ]   &    1  \\
000926   &2.07   &              &                 &     1.5e-6     [   0.26e-6  ] &    20-1.e4     &   4.73e52  [  1.3e52   ]&    310.  [ 20.   ]   &    2  \\
010222   &1.48   &              &                 &     5.7e-7     [   0.32e-7  ] &    20-1.e4     &   7.87e51  [  4.51e50  ]&    766   [ 30.   ]   &    2  \\
010921   &0.45   &-1.6  [ 0.1  ]&                 &     9.2e-7     [   1.4e-7   ] &    20-1.e4     &   7.33e50  [  1.33e50  ]&    129.  [ 26.   ]   &    2  \\
011211   &2.140  &-0.84 [ 0.09 ]&                 &     5.0e-8     [   1.e-8    ] &    40-700      &   3.17e51  [  0.32e51  ]&    185   [ 25    ]   &    1  \\
020124   &3.198  &-0.87 [ 0.17 ]&  -2.6   [ 0.65 ]&     9.4        [   1.8      ] &    2.-400      &   5.12e52  [  2.03e52  ]&    390   [ 113   ]   &    1  \\
020405   &0.695  &-0.0  [ 0.25 ]&  -1.87  [ 0.23 ]&     5.e-6      [   0.2e-6   ] &    15-2000     &   1.38e52  [  7.83e50  ]&    617   [ 171   ]   &    5  \\
020813   &1.255  &-1.05 [ 0.11 ]&                 &     32.3       [   2.1      ] &    2-400       &   2.58e52  [  2.4e51   ]&    478   [ 95    ]   &    1  \\
020819B  &0.41   &-0.9  [ 0.15 ]&  -2.0   [ 0.35 ]&     7.e-7      [   0.7e-7   ] &    25-100      &   1.49e51  [  3.23e50  ]&    70.   [ 21.   ]   &    7  \\
020903   &0.25   &-1.0  [ 0.0  ]&                 &     2.8        [   0.7      ] &    2-400       &   6.7e48   [  0.26e48  ]&    3.37  [ 1.79  ]   &    6  \\
021004   &2.335  &-1.0  [ 0.2  ]&                 &     2.7        [   0.5      ] &    2-400       &   4.6e51   [  0.12e51  ]&    267   [ 117   ]   &    6  \\
021211   &1.01   &-0.85 [ 0.09 ]&  -2.37  [ 0.42 ]&     30         [   2        ] &    2-400       &   7.13e51  [  9.9e50   ]&    94    [ 19    ]   &    1  \\
030226   &1.986  &-0.9  [ 0.2  ]&                 &     2.7        [   0.6      ] &    2-400       &   8.52e51  [  2.23e51  ]&    290   [ 63    ]   &    1  \\
030328   &1.520  &-1.14 [ 0.03 ]&  -2.1   [ 0.3  ]&     11.6       [   0.9      ] &    2-400       &   1.1e52   [  1.55e51  ]&    328   [ 35    ]   &    1  \\
030329   &0.169  &-1.32 [ 0.02 ]&  -2.44  [ 0.08 ]&     451        [   25       ] &    2-400       &   1.91e51  [  2.37e50  ]&    79    [ 3     ]   &    1  \\
030429   &2.656  &-1.1  [ 0.3  ]&                 &     3.8        [   0.8      ] &    2-400       &   7.6e51   [  1.47e51  ]&    128   [ 37    ]   &    1  \\
040924   &0.859  &-1.17  [0.05] &                 &     2.6e-6     [   0.3e-6   ] &    20-500      &   6.1e51   [  1.1e51   ]&    102   [ 35.   ]   &    1  \\
041006   &0.716  &-1.37 [ 0.14 ]&                 &     1.0e-6     [   0.1e-6   ] &    25-100      &   8.65e51  [  1.36e51  ]&    108   [ 22    ]   &    1  \\
050126   & 1.29  & -0.75 [0.44 ]&                 &   0.698        [  0.07    ]   &    15-150      &   1.12e51  [  0.25e51 ] &    263   [  110  ]   &  8    \\
050223   & 0.5915& -1.5  [0.42 ]&                 &   0.7          [  0.1     ]   &    15-150      &   1.43e50  [  0.2e50  ] &    110   [  54   ]   &  8    \\
050318   & 1.44  & -1.34 [0.32 ]&                 &   3.2          [  0.3     ]   &    15-150      &   5.11e51  [  0.8e51  ] &    115   [  27   ]   &  8    \\
050401   & 2.9   & -1.0  [0.0  ]& -2.45  [0.0  ]  &   2.45e-6      [  0.12e-6 ]   &    20-2000     &   2.03e53  [  0.1e53  ] &    501   [  117  ]   &  9    \\
050416A  & 0.653 & -1.01 [0.0  ]& -3.4   [0.0  ]  &   5.0          [  0.5     ]   &    15-150      &   9.3e50   [  0.9e50  ] &    28.6  [  8.3  ]   &  10   \\
050505   & 4.27  & -0.95 [0.31 ]&                 &   2.2          [  0.3     ]   &    15-350      &   5.65e52  [  0.8e52  ] &    661.  [  245  ]   &  11   \\
050525A  & 0.606 & -0.01 [0.11 ]&                 &   47.7         [  1.2     ]   &    15-350      &   9.53e51  [  2.5e51  ] &    127   [  5.5  ]   &  12   \\
050603   & 2.821 & -0.79 [0.06 ]& -2.15  [0.09 ]  &   3.2e-5       [  0.32e-5 ]   &    20-3000     &   2.13e54  [  0.22e54 ] &    1333  [  107  ]   &  13   \\
050803   & 0.422 & -0.99 [0.37 ]&                 &   1.5          [  0.2     ]   &    15-350      &   1.31e50  [  2.6e49  ] &    138   [  48   ]   &  14   \\
050814   & 5.3   & -0.58 [0.56 ]&                 &   1.0          [  0.3     ]   &    15-350      &   3.0e52   [  5.6e51  ] &    339   [  47   ]   &  15   \\
050820A  & 2.612 & -1.12 [0.14 ]&                 &   1.3e-6       [  0.13e-6 ]   &    20-1000     &   9.1e52   [  6.8e51  ] &    1325  [  277  ]   &  16   \\
050904   & 6.29  & -1.11 [0.06 ]& -2.2   [0.4  ]  &   0.8          [  0.2     ]   &    15-150      &   1.1e53   [  3.9e52  ] &    3178  [  1094.]   &  17   \\
050908   & 3.344 & -1.26 [0.48 ]&                 &   0.7          [  0.1     ]   &    15-150      &   8.29e51  [  1.3e51  ] &    195   [  36   ]   &  18   \\
050922C & 2.198  & -0.83  [0.26 ]&                 &   4.5e-6       [  0.7e-6  ]  &    20-2000 &       1.9e53   [  2.3e51  ] &    417  [  118  ]  &  19   \\
051022  & 0.80   & -1.176 [0.038]&                 &   1.e-5        [  0.13e-5 ]  &    20-2000 &       3.57e52  [  2.7e51  ] &    918  [  63   ]  &  20   \\
051109A & 2.346  & -1.25  [0.5  ]&                 &   5.8e-7       [  2.e-7   ]  &    20-500  &       3.87e52  [  3.8e51  ] &    539  [  381  ]  &  21   \\
060115  & 3.53   & -1.13  [0.32 ]&                 &   0.9          [  0.1     ]  &    15-150  &       1.24e52  [  2.0e51  ] &    288  [  47   ]  &  22   \\
060124  & 2.297  & -1.48  [0.02 ]&                 &   2.7e-6       [  0.8e-6  ]  &    20-2000 &       1.42e53  [  1.35e51 ] &    636  [  162  ]  &  23   \\
060206  & 4.048  & -1.06  [0.34 ]&                 &   2.8          [  0.2     ]  &    15-150  &       5.57e52  [  9.0e51  ] &    381  [  98   ]  &  24   \\
060210  & 3.91   & -1.12  [0.26 ]&                 &   2.7          [  0.3     ]  &    15-150  &       5.95e52  [  8.0e51  ] &    575  [  186  ]  &  25   \\
060218  & 0.0331 & -1.622 [0.16 ]&                 &   1.e-8        [  0.1e-8  ]  &    15-150  &       1.34e47  [  0.3e47  ] &    4.9  [  0.3  ]  &  26   \\
060223A & 4.41   & -1.16  [0.35 ]&                 &   1.4          [  0.2     ]  &    15-150  &       3.27e52  [  5.5e51  ] &    339  [  63   ]  &  27   \\
060418  & 1.489  & -1.5   [0.15 ]&                 &   6.7          [  0.4     ]  &    15-150  &       1.89e52  [  1.59e51 ] &    572  [  114  ]  &  28   \\
060510B & 4.9    & -1.53  [0.19 ]&                 &   0.6          [  0.1     ]  &    15-150  &       2.26e52  [  1.78e51 ] &    575  [  227  ]  &  29   \\
\hline
\end{tabular}
\caption{continue....}
\end{table*}
\setcounter{table}{2}
\begin{table*}
\begin{tabular}{lllllllllll} 
\hline 
GRB        &$z$    &$\alpha$        &$\beta$         &Peak Flux$^a$       &Range     &$L_{\rm iso}$  & $E_{\rm peak}$    & Ref \\ 
           &       &                &                &                    &keV       &erg/s            &keV              &     \\
\hline
060522  & 5.11   & -0.7   [0.44 ]&                 &   0.6          [  0.1     ]  &    15-150  &       2.0e53   [  3.7e51  ] &    427  [  79   ]  &  30   \\
060526  & 3.21   & -1.1   [0.4  ]& -2.2   [0.4  ]  &   1.7          [  0.1     ]  &    15-150  &       1.72e52  [  3.1e51  ] &    105.2[  21.1 ]  &  31   \\
060605  & 3.78   & -1.0   [0.44 ]&                 &   0.5          [  0.1     ]  &    15-150  &       9.5e51   [  1.5e51  ] &    490  [  251  ]  &  32   \\
060607A & 3.082  & -1.09  [0.19 ]&                 &   1.4          [  0.1     ]  &    15-150  &       2.0e52   [  2.7e51  ] &    575  [  200  ]  &  33   \\
060614  & 0.125  &               &                 &   11.6         [  0.7     ]  &    15-150  &       5.3e49   [  1.4e49  ] &    55   [  45   ]  &  34   \\
060707  & 3.43   & -0.73  [0.4  ]&                 &   1.1          [  0.2     ]  &    15-150  &       1.4e52   [  2.8e51  ] &    302  [  42   ]  &  35   \\
060714  & 2.711  & -1.77  [0.24 ]&                 &   1.4          [  0.1     ]  &    15-150  &       1.42e52  [  1.e51   ] &    234  [  109  ]  &  36   \\
060814  & 0.84   & -1.43  [0.16 ]&                 &   2.13e-6      [  0.35e-6 ]  &    20-1000 &       1.e52    [  1.e51   ] &    473  [  155  ]  &  37   \\
060904B & 0.703  & -1.07  [0.37 ]&                 &   2.5          [  0.1     ]  &    15-150  &       7.38e50  [  1.4e50  ] &    135  [  41   ]  &  38   \\
060906  & 3.686  & -1.6   [0.31 ]&                 &   2.0          [  0.3     ]  &    15-150  &       3.55e52  [  3.9e51  ] &    209  [  43   ]  &  39   \\
060908  & 2.43   & -0.9   [0.17 ]&                 &   3.2          [  0.2     ]  &    15-150  &       2.6e52   [  4.6e51  ] &    479  [  110  ]  &  40   \\
060927  &  5.6   & -0.93  [0.38 ]&                 &   2.8          [  0.2     ]  &    15-150  &       1.14e53  [  2.0e52  ] &    473  [  116  ]  &  41   \\
061007  & 1.261  & -0.7   [0.04 ]& -2.61  [0.21 ]  &   1.95e-5      [  0.28e-5 ]  &    20-1e4  &       1.74e53  [  2.45e52 ] &    902  [  43   ]  &  42   \\
061121  & 1.314  & -1.32  [0.05 ]&                 &   1.28e-5      [  0.17e-5 ]  &    20-5000 &       1.41e53  [  1.5e51  ] &    1289 [  153  ]  &  43   \\
061126  & 1.1588 & -1.06  [0.07 ]&                 &   9.8          [  0.4     ]  &    15-150  &       3.54e52  [  3.0e51  ] &    1337 [  410  ]  &  44   \\
061222B & 3.355  & -1.3   [0.37 ]&                 &   1.5          [  0.4     ]  &    15-150  &       1.82e52  [  2.75e51 ] &    200  [  28   ]  &  45   \\
070125  & 1.547  & -1.1   [0.1  ]& -2.08  [0.13 ]  &   2.25e-5      [  0.35e-5 ]  &    20-1.e4 &       3.24e53  [  5.e52   ] &    934  [  148  ]  &  46   \\
070508  & 0.82   & -0.81  [0.07 ]&                 &   8.3e-6       [  1.1e-6  ]  &    20-1000 &       3.3e52   [  3.9e51  ] &    342  [  15   ]  &  47   \\
071003  & 1.100  & -0.97  [0.07 ]&                 &   1.22e-5      [  0.2e-5  ]  &    20-4000 &       8.4e52   [  1.5e51  ] &    1678 [  231  ]  &  48   \\
071010B & 0.947  & -1.25  [0.6  ]& -2.65  [0.35 ]  &   8.92e-7      [  3.7e-7  ]  &    20-1000 &       6.4e51   [  5.3e49  ] &    101  [  23   ]  &  49   \\
071020  & 2.145  & -0.65  [0.3  ]&                 &   6.04e-6      [  2.1e-6  ]  &    20-2000 &       2.2e53   [  9.6e51  ] &    1013 [  205  ]  &  50   \\
071117  & 1.331  & -1.53  [0.15 ]&                 &   6.66e-6      [  1.8e-6  ]  &    20-1000 &       1.e53    [  7.e51   ] &    648  [  318  ]  &  51   \\
080319B & 0.937  & -0.82  [0.01 ]& -3.87  [0.8  ]  &   2.17e-5      [  0.21e-5 ]  &    20-7000 &       9.6e52   [  2.3e51  ] &    1261 [  25   ]  &  52   \\
080319C & 1.95   & -1.2   [0.1  ]&                 &   3.35e-6      [  0.74e-6 ]  &    20-4000 &       9.5e52   [  1.2e51  ] &    1752 [  505  ]  &  53   \\
\hline
\end{tabular}
\caption{
$^a$Peak Fluxes are in erg/s/cm$^2$ or photons/s/cm$^2$.
Reference for the Peak Flux (or Luminosity): 
1) Firmani et al. 2006;
2) Ulanov et al. 2005 ($L_{\rm iso}$ computed as $[P/F] [1+z] E_{\rm iso}$); 
3) Jimenez et al. 2001; 
4) Amati et al. 2002; 
5) Price et al. 2003; 
6) Sakamoto et al. 2005; 
7) Hurley et al. GCN 1507; 
8) GRB BAT on line table (http://swift.gsfc.nasa.gov/docs/swift/archive/grb\_table/);
9) Golenetskii et al. 2005f, GCN 3179; 
10) Sakamoto et al. 2006a 
11) Hullinger et al. 2005; 
12) Blustin et al. 2006; 
13) Golenetskii et al. 2005g; 
14) Parson et al. 2005; 
15) Tueller et al. 2005; 
16) Cenko et al., 2006; 
17) Sakamoto et al. 2005b; 
18) Sato et al. 2005; 
19) Golenetskii et al. 2005h; 
20) Golenetskii et al. 2005i; 
21) Golenetskii et al. 2005j; 
22) Barbier et al. 2006a; 
23) Golenetskii et al. 2006j; 
24) Palmer et al. 2006a; 
25) Sakamoto et al. 2006b; 
26) Campana et al., 2006;
27) Cummings et al. 2006a; 
28) Cummings et al. 2006b; 
29) Barthelmy et al. 2006; 
30) Krimm et al. 2006a; 
31) Markwardt et al. 2006a; 
32) Sato et al. 2006a; 
33) Tueller et al. 2006; 
34) Mangano et al. 2007; 
35) Stamatikos et al. 2006a; 
36) Krimm et al. 2006b;
37) Golenetskii et al. 2006k; 
38) Markwardt et al. 2006b; 
39) Sato et al. 2006b; 
40) Palmer et al. 2006b; 
41) Stamatikos et al. 2006b; 
42) Golenetskii et al. 2006l; 
43) Golenetskii et al. 2006m; 
44) Krimm et al. 2006c; 
45) Barbier et al. 2006b; 
46) Golenetskii et al. 2007l; 
47) Golenetskii et al. 2007m; 
48) Golenetskii et al. 2007n; 
49) Golenetskii et al. 2007o; 
50) Golenetskii et al. 2007p; 
51) Golenetskii et al. 2007q; 
52) Golenetskii et al. 2007r; 
53) Golenetskii et al. 2007s.}   
\label{tabz}
\end{table*}

\end{document}